\documentclass[11pt]{article}

\usepackage{authblk}

\usepackage[T1]{fontenc}
\usepackage{mathpazo}  

\usepackage{amsmath, amssymb, amsthm}
\usepackage{thmtools}

\usepackage[colorlinks=true,linkcolor=blue,citecolor=blue,urlcolor=blue]{hyperref}
\usepackage{cleveref}
\usepackage{bm}
\usepackage{enumitem}  
\usepackage[margin=1in]{geometry}  
\usepackage{comment} 
\usepackage[most]{tcolorbox}
\usepackage{multirow}

\newtheorem{theorem}{Theorem}[section]
\newtheorem{lemma}[theorem]{Lemma}
\newtheorem{proposition}[theorem]{Proposition}

\newtheorem{claim}[theorem]{Claim}

\theoremstyle{definition}
\newtheorem{definition}[theorem]{Definition}

\theoremstyle{remark}

\title{Coresets for Capacitated Clustering via Dual Concentration}
\author{
  \makebox[\textwidth][c]{Milind Prabhu \qquad Chris Schwiegelshohn \qquad Sudarshan Shyam \qquad Erik Waingarten}
}

\date{~}

\usepackage{algorithm}
\usepackage{algpseudocode}
\newcommand{\opt}{\mathrm{OPT}}
\newcommand{\eps}{\varepsilon}
\newcommand{\R}{\mathbb{R}}

\newcommand{\cost}{\mathrm{cost}}
\newcommand{\poly}{\mathrm{poly}}
\newcommand{\OPT}{\mathrm{OPT}}
\newcommand{\Var}{\operatorname{Var}}

\newcommand{\kmedian}{k\text{-}\mathrm{median}}
\newcommand{\kmeans}{k\text{-}\mathrm{means}}

\newcommand{\milind}[1]{\textcolor{blue}{[Milind: #1]}}

\newcommand{\TCost}{\widetilde{\operatorname{cost}}}
\newcommand{\DualCost}{\mathsf{Dual}}
\newcommand{\Error}{\mathsf{Err}}
\newcommand{\loadvector}{\text{demand vector }}
\newcommand{\loadvectors}{\text{demand vectors }}

\def\Allow(#1){\ensuremath{A_C(#1)}}

\newcommand{\Wone}{W_1}
\newcommand{\Wtwo}{W_2}

\newcommand{\pdim}{\operatorname{pdim}}
\newcommand{\Gap}{\mathcal G}
\begin{document}

\maketitle

\begin{abstract}
We study coresets for clustering problems, focusing on the classical
\(\kmedian\) and \(\kmeans\) objectives. A coreset is a small weighted
summary of the input data that approximately preserves the clustering
cost for every choice of centers. Such summaries are a central tool for
scaling clustering algorithms to large datasets.

In this work, we consider the capacitated setting, where the clustering
is subject not only to choosing centers but also to respecting prescribed
cluster-size constraints. This model is a natural generalization of
standard clustering and has been widely studied, but its coreset
complexity is less well understood. For finite metric spaces, we obtain
optimal-size coresets; surprisingly, these match the best-known bounds
for unconstrained clustering. We also give improved coreset bounds in Euclidean space, notably the first coresets with near-linear dependence in $k$ while being independent of the dimension and input size. 

The main new technical ingredients are concentration bounds for the dual solutions of min-cost flow problems, which allow for both significantly simpler and tighter analyses of the underlying sampling distribution used in the coreset construction. 
\end{abstract}

\paragraph{Statement on AI Use.}
The perspective of using the dual for concentration, rather than the primal, was pointed out by ChatGPT-5.5 Pro. In addition, the authors used AI to help polish the writing, review the proofs, and check the manuscript for possible errors, in multiple rounds of prompting. All content was carefully examined by the authors, who take full responsibility for the content of the manuscript.

\thispagestyle{empty} 
\clearpage 
\newpage

\pagenumbering{arabic}
\section{Introduction}
Coresets are a staple of big data algorithms. Given a loss function, a
point set \(P\), and a set of candidate solutions \(\mathcal C\), a coreset
\(Q\subseteq P\) is a significantly smaller weighted set such that, for all
\(C\in\mathcal C\),
\[
\cost(Q,C)=(1\pm\varepsilon)\cost(P,C).
\]
For a fixed accuracy parameter \(\varepsilon\), the quality of a coreset
algorithm is  measured by the number of distinct points in \(Q\), which we call its \emph{size}.  

In this paper, we study coresets for center-based clustering where we are typically interested in objectives of
the form \(\sum_{p\in P} d^z(p,C(p))\), where \(d\) is the distance
function of the underlying metric and \(C(p)\) is the center to which
\(p\) is assigned from a set \(C\) of at most \(k\) centers. The most
important special cases, \(z=1\) and \(z=2\), are known as \(\kmedian\)
and \(\kmeans\), respectively. For the unconstrained variants, where
\(C(p)=\arg \min_{c\in C}d(p,c)\), the coreset problem is now well
understood. Specifically, for finite \(N\)-point metrics, coresets of size
\(\widetilde O(k\varepsilon^{-2}\log N)\) are known
\cite{FL11,CSS21}; in Euclidean spaces, coresets of size
\(\widetilde O(k\varepsilon^{-2}\min(k^{z/(z+2)},\varepsilon^{-z}))\)
are known \cite{Cohen-AddadLSSS22,HLW23}. These bounds are optimal up to
logarithmic factors \cite{CGSS22,HLW23}. A more detailed review of
coreset bounds for different metric spaces appears in \Cref{sec:related}.

The situation is markedly different for constrained variants of these
problems. A central example is cardinality-constrained clustering. In
this paper we use the standard fractional transportation formulation:
given a demand vector \(h\in n\Delta_k\), the objective is to minimize
the \(\kmedian\) or \(\kmeans\) cost over assignments
\(x_{p,i}\in[0,1]\) such that each point \(p\in P\) is assigned total
mass \(\sum_{i=1}^k x_{p,i}=1\), and each center \(c_i\) receives total
mass \(\sum_{p\in P}x_{p,i}=h_i\). For unweighted point sets and integral demands, this coincides with the usual requirement that
exactly \(h_i\) points are assigned to \(c_i\).

Our coreset guarantee preserves this objective simultaneously for every
choice of centers \(C\) and every demand vector \(h\). Consequently, it
also preserves the optimum over any feasible family of demand vectors.
Via known reductions, see \Cref{prop:reduction}, this yields coresets
for widely studied constrained clustering problems, including upper- and
lower-capacitated clustering, fair clustering, and clustering with
anonymity constraints.

Having established that the same coreset framework applies to these
constrained variants, we next ask how large such coresets must be. More
precisely, we are interested in the dependence of the coreset size on the
main problem parameters, such as \(k\), the accuracy parameter
\(\varepsilon\), and the dimension or metric size. This dependence is
difficult to summarize succinctly. The most important barrier is that all
known constructions have at least quadratic dependence on \(k\)
\cite{BFS21,BravermanCJKST022,Cohen-AddadL19,Huang0L025}, in addition to
non-optimal dependencies on other parameters. Such a quadratic dependence
has historically been a major barrier even for unconstrained clustering;
for cardinality-constrained clustering, it has so far proven formidable.
At the same time, there is no evidence that cardinality constraints should
force larger coresets than in the unconstrained setting. Thus, the guiding
question of this paper is:

\begin{tcolorbox}[colback=gray!10,colframe=gray!40,boxrule=0.5pt,arc=2pt,
left=6pt,right=6pt,top=4pt,bottom=4pt]
\begin{quote}
Do cardinality constraints fundamentally change coreset sizes for clustering?
\end{quote}
\end{tcolorbox}



\subsection{Our Results} In this paper, we give strong evidence that the answer may be negative in general and resolve the question for finite metrics. Notably, our first main result is: \begin{theorem} There exists a coreset for $k$-median and $k$-means with cardinality constraints in finite $N$-point metrics of size $O(k\varepsilon^{-2}\log N)$. \end{theorem} 
This bound matches the best-known lower bounds for unconstrained
\(\kmedian\) and \(\kmeans\) in finite metrics~\cite{CGSS22}, and is
therefore optimal up to constant factors. 
Combining our finite-metric coreset with previous techniques for cardinality-constrained coreset construction already gives improved Euclidean bounds over prior work \cite{BravermanCJKST022,Huang0L025}; see \Cref{sec:overview}. However, this black-box route still has quadratic dependence on \(k\).
Our second main result breaks the quadratic barrier in \(k\), with no dependence on the dimension \(d\) or the input size $n$. 

\begin{theorem}
There exists a coreset for $k$-median $(z=1)$ and $k$-means $(z=2)$ with cardinality constraints in $d$-dimensional Euclidean spaces of size
\(
\widetilde O(k\varepsilon^{-2-2z}).
\)
\end{theorem}

This does not match the known lower bounds for Euclidean clustering \cite{HLW23}. Nevertheless, to the best of our knowledge, it is the first cardinality-constrained coreset construction to combine near-linear dependence on \(k\) with competitive \(\varepsilon\)-dependence. The algorithms are efficient with the  bottleneck being the computation of a constant-factor approximation.

\begin{table}[ht!]
\begin{center}
\begin{tabular}{r|c}
Reference & Coreset Size \((k\text{-median}, k\text{-means})\) 
\\\hline\hline
\multirow{2}{*}{Cohen-Addad, Li~\cite{Cohen-AddadL19}} & $\tilde{O}(k^2 \varepsilon^{-5}\log^2 n )$ \\
& $\tilde{O}(k^5 \varepsilon^{-5}\log^5 n )$ 
\\
\hline
\multirow{2}{*}{Schmidt, Schwiegelshohn, Sohler \cite{SSS19}} & $O(k \varepsilon^{-d-1}\log n)$
\\
& $O(k \varepsilon^{-d-2}\log n)$ \\
\hline
\multirow{2}{*}{Huang, Jiang, Vishnoi~\cite{HuangJV19}} & $O(k^2 \varepsilon^{-d})$\\
& $O(k^3\varepsilon^{-d-1})$
\\
\hline
Braverman, Cohen-Addad, Jiang, Krauthgamer, &  $\tilde{O}(k^3 \varepsilon^{-6})$ 
\\
Schwiegelshohn, Toftrup, Wu~\cite{BravermanCJKST022} & $\tilde{O}(k^7 \varepsilon^{-10})$\\
\hline
\multirow{2}{*}{Huang, Li, Lu, Wu~\cite{Huang0L025}}& $\tilde{O}(k^2 \varepsilon^{-4})$  
\\
& $\tilde{O}(k^2 \varepsilon^{-6})$
\\
\hline
\multirow{2}{*}{\textbf{This paper}} & $\tilde{O}(k \varepsilon^{-4})$ \\
& $\tilde{O}(k \varepsilon^{-6})$
\end{tabular}
\end{center}
\caption{Coreset size bounds for cardinality-constrained clustering in $d$-dimensional Euclidean space. For each reference, the first bound is for \(k\)-median and the second
is for \(k\)-means.
}
\label{table}
\end{table}

\section{Preliminaries and Definitions}
\label{sec:problem-setup}

\paragraph{Notation.}
We use \(a\lesssim b\) to denote \(a\le \eta b\) for some absolute
constant \(\eta\). For an integer \(m\), write \([m]=\{1,\ldots,m\}\).
Let
\(
\Delta_k=\{x\in\R^k:x\ge 0,\ \sum_i x_i=1\}
\)
denote the \(k\)-dimensional unit simplex.

\paragraph{The \(\kmedian\) Objective with Cardinality Constraints.}
Let \((X,d)\) be a metric space. We first define the objective for a
generic weighted instance \(S\subseteq X\) of total weight \(n\). We write
\(w_S(s)\) for the weight of \(s\in S\) and $n = \sum_{s\in S} w_S(s)$ to denote the total weight of the instance.

Fix an ordered center tuple \(C=(c_1,\ldots,c_k)\in X^k\) and a
\loadvector \(h\in n\Delta_k\). The coordinate \(h_i\) is the demand of center \(c_i\). For this \loadvector \(h\), the
cardinality-constrained assignment cost, denoted by  \(\cost_S(C,h)\), is the cost of the cheapest fractional assignment of points in $S$ to the centers in $C$ that respects the demands $h$. Formally, it is the value of
the following LP.

\begin{center}
\begin{minipage}[t]{0.49\textwidth}
\begin{tcolorbox}[
    title=\textbf{Primal assignment LP},
    colback=white,
    colframe=black,
    colbacktitle=black,
    coltitle=white,
    boxrule=0.5pt,
    arc=1pt,
    left=3pt,
    right=3pt,
    top=4pt,
    bottom=4pt,
    height=4.35cm,
    valign=center,
    fontupper=\small
]
\[
\begin{aligned}
\cost_S(C,h)
=
\min_x\quad
&\,\sum_{s,i}x_{s,i}\,d(s,c_i)
\\
\mathrm{s.t.}\quad
&\sum_{i\in[k]}x_{s,i}=w_S(s)
\quad s\in S,
\\
&\sum_{s\in S}x_{s,i}=h_i
\quad i\in[k],
\\
&x_{s,i}\ge 0
\quad s\in S,\ i\in[k].
\end{aligned}
\]
\end{tcolorbox}
\end{minipage}
\hfill
\begin{minipage}[t]{0.49\textwidth}
\begin{tcolorbox}[
    title=\textbf{Dual assignment LP},
    colback=white,
    colframe=black,
    colbacktitle=black,
    coltitle=white,
    boxrule=0.5pt,
    arc=1pt,
    left=3pt,
    right=3pt,
    top=4pt,
    bottom=4pt,
    height=4.35cm,
    valign=center,
    fontupper=\small
]
\[
\begin{aligned}
\max_{\alpha\in\R^S,\ \beta\in\R^k}\quad
&\sum_{s\in S}w_S(s)\alpha_s-h\cdot\beta
\\
\mathrm{s.t.}\quad
&\alpha_s\le d(s,c_i)+\beta_i
\quad s\in S,\ i\in[k].
\end{aligned}
\]
\end{tcolorbox}
\end{minipage}
\end{center}

The dual of the above LP, written alongside, will be important in the analysis later in the paper. The key reason for this is that it can be expressed succinctly.  In particular, we can eliminate the auxiliary variables
\(\alpha_s\). For a fixed \(\beta\in\R^k\), the optimal choice of
\(\alpha_s\) is
\[
f_{C,\beta}(s)
=
\min_{i\in[k]}\{d(s,c_i)+\beta_i\}.
\]
Thus the reduced dual objective is
\begin{equation}
\label{eq:finite-dual-objective}
\DualCost_S(C,h,\beta)
=
\sum_{s\in S}w_S(s)f_{C,\beta}(s)-h\cdot\beta .
\end{equation}

We now note the following consequence of strong duality.
\begin{restatable}[Strong duality]{claim}{StrongDuality}
\label{lem:finite-strong-duality}
For every weighted instance \(S\subseteq X\) of total weight \(n\), every
center tuple \(C=(c_1,\ldots,c_k)\in X^k\), and every \loadvector
\(h\in n\Delta_k\),
\[
\cost_S(C,h)
=
\sup_{\beta\in\R^k}\DualCost_S(C,h,\beta).
\]
\end{restatable}

The same notation applies to \(\kmeans\), with each assignment cost
\(d(s,c_i)\) replaced by \(d(s,c_i)^2\). The two ambient metric spaces we
consider are finite metric spaces, where \(X\) is finite and \(N=|X|\),
and Euclidean spaces, where \(X=\R^d\) and \(d(s,c)=\|s-c\|\). Thus, in
the Euclidean sections, the \(\kmedian\) and \(\kmeans\) assignment costs
are \(\|s-c_i\|\) and \(\|s-c_i\|^2\), respectively.

\paragraph{Coresets.}
 The input is an unweighted finite
set \(P\subseteq X\), and we write \(n=|P|\). 

\begin{definition}[\(\eps\)-coreset]
For $\eps \in (0,1)$, a weighted set \(Q\subseteq X\) of total weight \(n\) is an
\(\eps\)-coreset for \(P\) if, for every center tuple
\(C=(c_1,\ldots,c_k)\in X^k\) and every \loadvector \(h\in n\Delta_k\),
\[
\left|
\cost_Q(C,h)-\cost_P(C,h)
\right|
\le
\eps\cdot \cost_P(C,h).
\]
The \emph{size} of \(Q\) is the number of distinct points in its support.
\end{definition}

For convenience of exposition, we assume throughout that the input support
has size at least \(k\eps^{-1}\). Otherwise, storing the entire input
support is already within the claimed coreset size bounds. In the
finite-metric setting, this assumption implies \(N\ge k\eps^{-1}\), and
hence
\(
\log(Nk/\eps)\le 2\log N .
\)
Thus, in the finite-metric bounds, factors of \(\log(Nk/\eps)\) are
absorbed into \(\log N\).

\section{Technical Overview}
\label{sec:overview}

In this section, we outline the key ideas underlying our analysis. We begin with the problem that admits the simplest analysis: capacitated \(k\)-median in finite metrics. Both capacitated $k$-means in finite metrics as well as the Euclidean space constructions follow the same basic outline, but require significant extensions.

\subsection{Finite-Metric Capacitated \(\kmedian\)}

Roughly speaking, the best coreset constructions have their technical origins in Chen's seminal ring-sampling algorithm~\cite{Chen09}, which we also use and proceeds by: (i) finding a distribution supported on the $n$ input points $P$ such that for any (fixed) set of candidate centers $C$, a small re-weighted sample from the distribution provides an accurate estimate for the cost of clustering $P$ with $C$; then, (ii) taking enough samples from the distribution so as to union-bound over all candidate centers $C$. The re-weighted set of samples drawn becomes the coreset.

Perhaps implicit, one conceptual core of Chen's framework is that, for any fixed candidate set of centers $C$, the (unconstrained) clustering cost of $P$ with $C$ is a sum of $n$ ``local'' quantities, where each point of \(P\) contributes its distance to the closest center in $C$, and  estimating large sums via random sampling is, at least by now, a well-established algorithmic technique. In capacitated clustering, this decomposition of cost via $n$ ``local'' summands  breaks down.
\begin{quote}
\textbf{Main Challenge of Capacitated Clustering}. In capacitated clustering, the optimal clustering cost of $P$ with respect to a fixed set of centers $C$ is not a sum of individual costs over $n$ dataset points; rather, it is the result of a linear program whose point-to-center assignments must satisfy global constraints (the capacities of each center).
\end{quote}
This nonlocality is a central obstacle in capacitated clustering
coresets~\cite{Cohen-AddadL19,BravermanCJKST022,Huang0L025}, and is
already visible in the following simple example. Suppose $P$ consists of two far-away clusters of $n/2$ points and that $C$ consists of two centers, $c_1$ and $c_2$, one for each cluster with capacity $n/2$. If a coreset \(Q\) underrepresents one of the two
clusters, then any feasible assignment for \(Q\) must send some weight
across the large inter-cluster distance. Thus, a small error in the mass
represented inside one cluster can create a large error in the capacitated
clustering cost.


In the Euclidean setting, previous capacitated coresets handle this
problem by breaking the input into rings around a reference solution. They sample points uniformly inside each ring and give the sampled points weights so
that each ring has the correct total weight. This makes the sampling step
easier to analyze. However, the capacity constraints are still global: a
feasible assignment must send the right amount of mass to every center.
Prior analyses handle this by controlling the assignment error separately
inside the local pieces.

This local control is costly. Up to lower-order factors, Cohen-Addad and Li~\cite{Cohen-AddadL19} obtained a bound with
quadratic dependence on \(k\) and polylogarithmic dependence on \(n\)
for \(k\)-median. Braverman et
al.~\cite{BravermanCJKST022} removed the input-size dependence, but their
bound has a larger \(k^3\)-type dependence. Huang et
al.~\cite{Huang0L025} used a global sampling distribution together with
the preprocessing of~\cite{BravermanCJKST022}, and obtained a
\(k^2\)-type bound. Still, the coreset size remains quadratic in \(k\). We remove this
quadratic dependence by preserving the dual of the constrained assignment
problem directly.

\newcommand{\ignore}[1]{}

\ignore{\subsection{Capacitated $k$-Median in Finite Metrics}

Starting point for all works on coresets for capacitated clustering is Chen's seminal ring sampling algorithm \cite{Chen09}.
Chen reduced the problem of constructing a coreset from arbitrary instances to ring instances, that is the points can be assumed to have, up to constant factors, the same cost in an optimal or near-optimal solution. Uniform sampling is then the choice algorithm for constructing coresets for these rings. 

Cohen-Addad and Li \cite{Cohen-AddadL19} applied Chen's framework to capacitated coresets. Even though more performative sampling distributions for unconstrained clustering existed, uniform sampling was, up to that point, the only sampling distribution used in coreset construction that guaranteed mass preservation, that is the total weight of points in the coreset equals the total number of points in the original point set.

The necessity for this requirement already illustrates several difficulties when constructing coresets for capacitated clustering that are not present in unconstrained clustering.
If the total weight of points in the coreset $Q$ is not equal to the total number of points in the original point set $P$, then capacity constraints that are feasible for $Q$ might not be feasible for $P$ and vice versa. 
In fact, the same must be true for certain subsets of $Q$ and $P$; if there exist two clusters $C_1$ and $C_2$ that are very well separated, then the mass of $C_1\cap Q$ and $C_1\cap P$ (and respectively for $C_2$) must be preserved, as otherwise at least one point in say $Q$ might be forced to traverse the distance between $C_1$ and $C_2$, whereas the same might not be true for $P$.

In a nutshell, Chen's ring-sampling algorithm allowed Cohen-Addad and Li to localize the cost analysis, despite the inherently global nature of capacity constraints.
Unfortunately, Chen's reduction comes at a significant cost. The preprocessing leads to $k \log n$ rings, which automatically incurs the number of rings as an additional multiplicative overhead in the coreset size. With significant effort, \cite{BravermanCJKST022} managed to reduce the number of rings to $k^2 \varepsilon^{-z}$, where $z=1$ for $k$-median and $z=2$ for $k$-means. No coreset construction for capacitated clustering is known to avoid an inherent $k^2$ dependency in the coreset size \footnote{The only exception is the algorithm by \cite{SSS19} which has an exponential dependence on the dimension. While dimension reduction is possible, it would lead to a bound of the order $k^{\text{poly}(\varepsilon^{-1})}$ or even higher.}.}

\subsection{The Product Ring Sampling Algorithm}

Our algorithm for $k$-median in finite metrics also uses Chen's ring-decomposition, and blends it with a global sampling distribution. The algorithm follows the following steps:
\begin{enumerate}
    \item Compute an $O(1)$ approximate $k$-median clustering with center set $A$.
    \item Partition each cluster into $\log n$ rings after which any two points in the same ring will have up to constant factors the same cost in $A$.
    \item For a sampling budget $|Q|$, sample $m_R=\left\lceil |Q| \cdot \frac{\text{cost}(R,A)}{\text{cost}(P,A)}\right\rceil
$ points from ring $R$ and weigh each sampled point $q\in R\cap Q$ by $w_q =|R| / m_R$.
\end{enumerate}

In other words, as in Chen's algorithm, the sampling distribution is a
product of uniform distributions over the \(O(k\log n)\) rings. The
difference is that the sample size of a ring is chosen according to its
importance, namely its contribution to \(\cost(P,A)\). Because of the
ceilings, the total number of sampled points is at most
\(|Q|+O(k\log n)\). Importantly, the algorithm does not require the
sample from each ring to be a coreset for that ring by itself; such a
ring-by-ring guarantee would reintroduce the factor-\(k\) overhead
from~\cite{Cohen-AddadL19}. Instead, our analysis treats the samples from
all rings together, in one shot. A similar algorithm was proposed by
Huang et al.~\cite{Huang0L025}, using the more involved preprocessing
of~\cite{BravermanCJKST022} in place of Chen's decomposition.





\subsection{Assignment Cost Preservation via Dual Concentration}
Consider the dual program given in Equation \ref{eq:finite-dual-objective}. 
First note that $\mathbb{E}\left[\DualCost_Q(C,h,\beta)\right] = \DualCost_P(C,h,\beta)$. Thus, if we show that the value of every dual solution \(\beta\) concentrates around its expectation, then strong duality implies that \(\cost_Q(C,h)\) concentrates around \(\cost_P(C,h)\). 

At a high level, for a fixed solution \(C\) and a reference solution
\(A=\{a_1,\ldots,a_k\}\), we replace the pointwise dual contribution
\(
f_{C,\beta}(p) = \min_{i\in[k]}\{\operatorname{cost}(p,c_i)+\beta_i\}
\)
by its anchored version
\( g_{C,\beta}(p) = f_{C,\beta}(p)-f_{C,\beta}(a_j) \)
for \(p\in P_j\). The anchor contribution is preserved exactly by weight preservation within each region \(P_j\), so concentration is only needed for the lower-variance anchored terms.

Therefore, it suffices to prove concentration for $\sum_{q\in Q}w_q g_{C,\beta}(q)$. The main benefit of anchoring is that
$|g_{C,\beta}(q)|\leq d(q,A)$. This can be thought of a dual analogue to the variance reduction trick, initially introduced by \cite{CSS21} and subsequently used in most coreset analyses.
Since we can tolerate an error of the order $\varepsilon\sum_{p\in P}d(p,A)$, the anchoring technique effectively reduces the variance to a constant. Therefore, the modified anchored dual is preserved for any candidate $\beta$ by setting $|Q| \gtrsim \varepsilon^{-2}$.
What remains to be shown is that we can efficiently enumerate over the function space of anchored dual solutions, which, while non-trivial, is completely dominated by $N^k$ candidate solutions. Thus, setting $|Q| \gtrsim \varepsilon^{-2} k \log N$ yields a coreset for capacitated $k$-median.

\subsection{Capacitated $\kmeans$ in Finite Metrics}

For the $k$-means objective, we initially follow the same outline as $k$-median, that is, the algorithm is, save for the differences in computing an initial solution, identical.
The biggest difference is that the $k$-means objective requires significantly more effort to use the anchoring technique.
To highlight the difference, notice that
$g_{C,\beta}(p) = \min_{i\in[k]}\{\operatorname{cost}(p,c_i)+\beta_i\} - \min_{i\in[k]}\{\operatorname{cost}(a_j,c_i)+\beta_i\}$ for any $p$ assigned to $a_j$. Setting aside the $\beta_i$ values, we have $|d^2(p,c_i) - d^2(a_j,c_i)| \leq d(p,a_j)(d(p,c_i) + d(a_j,c_i))$, which is in general not of the order $\operatorname{cost}(p,a_j)$.
Indeed, without further modifications, it is difficult to provide any concentration.
We overcome this issue in two steps. First, we truncate the solution space, showing that we can reduce to the case $d(p,c_i)\leq \varepsilon^{-1}d(p,a_j)$. 
Second, we show that a weak concentration applying only to dual solutions that are near-optimal is sufficient and that achieving this weak concentration only requires $\varepsilon^{-2}k\log N$ many samples.

\paragraph{Solution Truncation.}
A commonly seen argument in coreset analyses for unconstrained clustering is that if one point $p$ of a ring is served at distance $\varepsilon^{-1}d(p,a_j)$, then all points in the same ring must be served at similar distance, and in fact, have nearly indistinguishable cost.
While this exact  property does not hold for capacitated clustering, one can nevertheless mimic features of this argument. Suppose that for some given solution $C$ and demand vector $h$, some fraction $f\cdot |R|$ of points from a ring $R$ will be served at distance at least $\varepsilon^{-1}d(p,a_j)$. In this case, we can consider the modified instance assigning the center $a_j$ instead of $p$, as the difference in cost is negligible. For the points where the distance depends on $p$, we thereafter have $g_{C,\beta}(p) \leq \varepsilon^{-1}\operatorname{cost}(p,a_j)$. Using Hoeffding's inequality shows that $|Q|\gtrsim \varepsilon^{-4}$ is sufficient to prove concentration for any fixed $\beta$. Thus, a reduction to truncated solutions already yields a coreset for $|Q|\gtrsim k \varepsilon^{-4}\log n$.

\paragraph{Weak Dual Convergence.}
For $k$-median, we simply preserved the cost of all dual solutions $\beta$ up to an additive error of $\varepsilon\cdot \sum_{p}d(p,A)$. This turns out to be overkill.
In order to conclude that the primal assignment cost is preserved under the coreset, we require only (i) that the cost of every near-optimal feasible dual solution $\beta$ is preserved and (ii) that the cost of $\beta$ that are far from optimal are not significantly overestimated. 
Similar ideas are sometimes used in the analysis of weak coresets, where preserving the cost of near-optimal solutions (provably) requires fewer samples than preserving the cost of all solutions \cite{prabhu2026approximation}. 



For a given set of centers \(C\) and demand vector \(h\), let \(\beta^*\) be an
optimal dual solution on \(P\), and consider any other dual solution \(\beta\).
If, for a sufficiently small absolute constant \(\eta\), the estimation error
for \(\DualCost_Q(C,h,\beta)\) is at most
\(
\eta\bigl(\DualCost_P(C,h,\beta^*)-\DualCost_P(C,h,\beta)\bigr),
\)
then \(\beta\) cannot attain a larger dual value than \(\beta^*\) on \(Q\).
In other words, the larger the gap between the dual cost of \(\beta^*\) and
\(\beta\) is, the coarser we can afford to be in our estimation of
\(\DualCost_Q(C,h,\beta)\).

\subsection{Euclidean Spaces}
A natural first attempt is to use the product-ring-sampling algorithm in
Euclidean space. The characterization of the solution space of candidate centers and dual vectors does not pose a significant bottleneck, with essentially a union bound over $\exp(\tilde{O}(kd))$ solutions being sufficient. Following a suitable dimension reduction, this may be further improved to $\exp(\tilde{O}(k\varepsilon^{-2}))$ many solutions. The problem with this approach is that we are guaranteed to sample at least one point from every ring, thus the overall coreset will have size $\tilde{O}(k\log n + k \varepsilon^{-4})$ for both $k$-median and $k$-means.
Using the ring-compression ideas from \cite{BravermanCJKST022} (see Theorem 3.2 of that reference), we can obtain a coreset of size $\tilde{O}(k^2\varepsilon^{-z} + k \varepsilon^{-4})$ in an almost black-box manner, where $z=1$ for $k$-median and $z=2$ for $k$-means\footnote{We do not need the full strength of the theorem. We only use
the preprocessing step to reduce the number of rings passed to the
product-ring sampler. See also \cite{Huang0L025}.}. Combining these algorithms will not avoid either the $k\log n$ or $k^2\varepsilon^{-z}$ terms, so to improve further, we require a new algorithm.

\paragraph{Mass-Preserving Sensitivity Sampling.}
Sensitivity sampling \cite{FL11,sensitivitycoresets}, which samples points according to their importance, is arguably the most widely used algorithmic framework in coreset construction. A popular implementation first computes a constant factor approximate solution $A$ of $P$ with  $P_j$ denoting the cluster induced by center $a_j$. Next, each point $p$ is sampled with probability proportional to $\frac{\text{cost}(p,A)}{\text{cost}(P,A)} + \frac{1}{|P_j|}$ and weighted appropriately. 
Notably, sensitivity sampling bypasses any need for ring construction, thus  avoiding the overhead in coreset size arising from the number of rings.

However, standard sensitivity sampling does not preserve mass exactly,
which is necessary for capacitated coresets. Previous works therefore
relied on uniform sampling, a natural choice when exact mass preservation
is required. Our approach instead starts with sensitivity sampling and
then performs a cluster-wise weight correction. Let \(P_j\) be the
reference cluster induced by anchor \(a_j\), and let \(Q_j\) be the
multiset of samples drawn from \(P_j\). Each sampled occurrence
\(q\in Q_j\) first receives its usual inverse-probability weight
\(\widehat w_q\). We then multiply all weights inside \(Q_j\) by the same
factor \(\lambda_j=|P_j|/\sum_{q\in Q_j}\widehat w_q\), and set
\(w_q=\lambda_j\widehat w_q\). Thus the final weights satisfy
\(\sum_{q\in Q_j}w_q=|P_j|\) exactly for every reference cluster \(P_j\).
This lets us retain the advantages of sensitivity-based sampling while
restoring the mass preservation needed in the capacitated setting.

\paragraph{Analysis.}
As a first observation, one might be tempted to reason that since, with very high probability $w_q$ and $\widehat w_q $ are very similar for all $q$, this modification does not significantly affect concentration properties.
This intuition is incomplete.
Before rescaling, the weighted sample need not preserve the cluster masses
exactly. Hence capacity constraints that are feasible for \(P\), and for
\(Q\) with final weights \(w_q\), need not be feasible for \(Q\) with
initial weights \(\widehat w_q\). Small changes in mass can therefore
have a large effect on feasibility, and consequently on cost.

The dual viewpoint is again useful here. Given centers \(C\) and a load
vector \(h\), our goal is to control the dual sampling error uniformly
over all dual vectors \(\beta\in\R^k\). Following the analysis of the
product ring sampling algorithm, we first use the fact that the term
\(-h\cdot\beta\) cancels when comparing \(\DualCost_Q(C,h,\beta)\) and
\(\DualCost_P(C,h,\beta)\). We then apply the anchoring trick and reduce the analysis of
\(\sum_{q\in Q}w_q g_{C,\beta}(q)\) to two estimates. We do this by
showing that, for all \(\beta\),
\begin{equation}
\label{eq:biaserror}
\left|
\sum_{q\in Q}(\widehat w_q-w_q)g_{C,\beta}(q)
\right|
\le
\varepsilon \sum_{p\in P} d(p,A)
\end{equation}
and
\begin{equation}
\label{eq:trueconcentration}
\left|
\sum_{q\in Q}\widehat w_q g_{C,\beta}(q)
-
\sum_{p\in P}g_{C,\beta}(p)
\right|
\le
\varepsilon \sum_{p\in P} d(p,A).
\end{equation}
The second estimate is a raw sensitivity-sampling concentration bound; it
does not use the final mass-preserving rescaling. The first estimate
controls the additional error caused by rescaling, using that
\(|w_q-\widehat w_q|\) is small and that
\(|g_{C,\beta}(q)|\le d(q,A)\).

One might ask where in this analysis we bypass the issue that some
\loadvectors may be feasible for \(Q\) after weight manipulation, but not
for the initially weighted sample. The crucial trick is that, after weight
manipulation, the anchor terms cancel:
\(\sum_{q\in Q_j} w_q f_{C,\beta}(a_j)
=
\sum_{p\in P_j} f_{C,\beta}(a_j)\).
In contrast,
\(\sum_{q\in Q_j} \widehat w_q f_{C,\beta}(a_j)
\neq
\sum_{p\in P_j} f_{C,\beta}(a_j)\)
in general, and the lack of anchor cancellation can lead to a large
difference between the true dual cost and the estimated dual cost with
initial weights \(\widehat w\). In summary, in addition to reducing the
variance, the anchoring trick isolates exactly where the weight
manipulation affects the Euclidean analysis. However, the trick is only
useful after the sample satisfies cluster-wise mass preservation. Without
mass preservation, convergence for
\(\sum_{q\in Q}w_q g_{C,\beta}(q)\) does not imply convergence for
\(\DualCost_Q(C,h,\beta)\).

As a final remark, we observe that the analysis for $k$-median is now effectively identical to that of finite metrics, with the only difference being in the enumeration over candidate solutions in Euclidean spaces. The underlying ideas for enumeration are largely a combination of enumerating solutions for unconstrained $k$-median clustering in Euclidean spaces, combined with dual concentration for all demand vectors analogously to the finite metric case.
For \(\kmeans\), combining mass-preserving sensitivity sampling with the
weak-dual-convergence argument from the finite-metric analysis appears
more delicate. We believe this is the main remaining step toward the
optimal \(\widetilde O(k\varepsilon^{-4})\) Euclidean \(\kmeans\) bound.

\section{Finite-Metric Capacitated $\kmedian$}
\label{sec:finite-kmedian}

We first prove the coreset theorem for capacitated $\kmedian$ in finite metrics.

\begin{theorem}[Finite-metric fractional capacitated $\kmedian$]
\label{thm:kmedian}
For every $\eps,\delta\in(0,1)$, there is a randomized algorithm which
outputs a weighted subset $Q\subseteq P$ of size
\[
|Q|
=
O\left(
\frac{k\log N +\log(1/\delta)}{\eps^2}
\right)
\]
such that, with probability at least $1-\delta$, $Q$ is an
$\eps$-coreset for finite-metric capacitated $\kmedian$.
\end{theorem}


We begin by giving the algorithm. As mentioned earlier, our algorithm is  similar to the one proposed by \cite{Huang0L025}. It first computes an unconstrained
reference $\kmedian$ solution. It then partitions the points into rings
around the reference centers. Finally, it samples points independently
and uniformly from each ring, with the number of samples from a ring
proportional to its contribution to the reference cost.
\paragraph{Coreset Construction.}
Let \(A=\{a_1,\ldots,a_k\}\subseteq X\) be a constant-factor approximate
unconstrained \(\kmedian\) solution for \(P\).  The points of \(A\) are
called anchors.  Assign each point \(p\in P\) to its nearest anchor,
breaking ties arbitrarily.  For an anchor \(a_j\), let \(P_j\) be its
assigned cluster, and
\[
\Phi_j=\sum_{p\in P_j} d(p,a_j).
\]
Then \(\sum_{j\in [k]}\Phi_j \lesssim \opt\), where \(\opt\) is the optimum
unconstrained \(\kmedian\) cost of \(P\).

We assume without loss of generality that no cluster has zero reference
cost. Indeed, if \(\Phi_j=0\), then all points in the cluster coincide with the
anchor, so we store a single point with the appropriate weight. This
represents the cluster exactly and increases the coreset size by at most \(k\).

\paragraph{Ring Partitioning.}
For each anchor \(a_j\), set
\(
\lambda_j=\frac{\Phi_j}{|P_j|}
\)
, and define the innermost ring by
\[
R_{j,0}
=
\{p\in P_j : d(p,a_j)\le \lambda_j\},
\]
and for \(t\ge 1\), define
\[
R_{j,t}
=
\{p\in P_j : 2^{t-1}\lambda_j < d(p,a_j) \le 2^t\lambda_j \}.
\]
Empty rings are discarded. Let \(\mathcal R\) denote the collection of all nonempty rings. We call these the sampled rings.
For a sampled ring \(R=R_{j,t}\), with anchor \(a_j\), write
\[
a_R=a_j,
\qquad
n_R=|R|,
\qquad
\rho_R=2^t\lambda_j .
\]
Thus,
\[
d(p,a_R)\le \rho_R
\qquad
\forall p\in R.
\]
The quantity \(n_R\rho_R\) is an upper bound, up to constants, on the total
anchor cost scale of ring \(R\). We use these quantities as the sampling
priorities of the rings. We normalize these scales by setting
\[
\Wone
=
\sum_{R\in\mathcal R} n_R\rho_R .
\]
The algorithm will allocate samples to each ring \(R\) proportionally to
\(n_R\rho_R/\Wone\).

Each anchor contributes at most \(O(\log n)\) nonempty
rings. Indeed, if the \(t\)-th ring around \(a_j\) is nonempty, then some
point in the ring has cost larger than \(2^{t-1}\lambda_j\). Since the
total cost of the cluster is \(\Phi_j=|P_j|\lambda_j\), we have
\(2^{t-1}\lambda_j<\Phi_j\), and hence \(t=O(\log |P_j|) = O(\log n)\). Therefore
\(|\mathcal R|=O(k\log n)\).

\begin{algorithm}[H]
\caption{Product Ring Sampling for Finite-Metric Capacitated \(\kmedian\)}
\label{alg:finite-kmedian-ring-sampling}
\begin{algorithmic}[1]
\State Construct the ring partition.
\State Set
\[
T=\Theta\left(
\frac{k\log N+\log(1/\delta)}{\eps^2}
\right).
\]
\State For every ring \(R\in\mathcal R\), sample
\[
m_R=\left\lceil T\frac{n_R\rho_R}{\Wone}\right\rceil
\]
points independently and uniformly from \(R\), where \(\Wone=\sum_{R\in\mathcal R} n_R\rho_R\)
\State Give each sampled point from \(R\) weight \(n_R/m_R\).
\State \textbf{Output:} the resulting weighted multiset \(Q\).
\end{algorithmic}
\end{algorithm}

\noindent
The total number of sampled points is \(O(T)\), because
\[
\sum_{R\in\mathcal R}m_R
\le
|\mathcal R|+T
=
O(T).
\]

\noindent
The following claim shows that the normalization constant $\Wone$ is controlled by
the reference cost, and hence by \(\opt\).

\begin{claim}
\label{lem:finite-kmedian-ring-scale}
 \(\Wone =\sum_{R\in\mathcal R}n_R\rho_R  \lesssim \opt\).
\end{claim}

\begin{proof}
Fix an anchor \(a_j\).  The innermost
ring has scale \(\lambda_j\), so its contribution is at most
\(|P_j|\lambda_j=\Phi_j\).  For every outer ring \(R\subseteq P_j\), each
point \(p\in R\) satisfies \(d(p,a_R)>\rho_R/2\).  Therefore
\(
n_R\rho_R
\le
2\sum_{p\in R}d(p,a_R).
\)
Summing first over the rings of \(P_j\), and then over all anchors yields the claim.
\end{proof}



\subsection{Analysis}\label{sec:finite-kemedian-analysis}
We now begin by recalling some of the notation introduced earlier.
Fix an ordered center tuple \(C=(c_1,\ldots,c_k)\in X^k\). 
For \(\beta\in\mathbb R^k\), recall
\[
f_{C,\beta}(p)
=
\min_{i\in[k]}\{d(p,c_i)+\beta_i\}.
\]
And for a \loadvector \(h\in n\Delta_k\),
\begin{align}
\DualCost_P(C,h,\beta)
=
\sum_{p\in P} f_{C,\beta}(p)-h\cdot\beta, \qquad
\DualCost_Q(C,h,\beta) = \sum_{q\in Q} w_q f_{C,\beta}(q)-h\cdot\beta .
\end{align}

\noindent We will use \Cref{lem:finite-strong-duality} (strong duality) throughout to pass between
primal costs and the corresponding dual objectives.

\noindent For fixed \(C\) and \(\beta\), the difference between the two dual
objectives does not depend on the demand vector, because the terms
\(-h\cdot\beta\) cancel. The dual sampling error is given by
\begin{equation}
\label{eq:kmedian-dual-sampling-error}
\Error_C(\beta)
=
\sum_{q\in Q}w_q f_{C,\beta}(q)
-
\sum_{p\in P}f_{C,\beta}(p).
\end{equation}
Equivalently, for every \(h\in n\Delta_k\),
\[
\Error_C(\beta)
=
\DualCost_Q(C,h,\beta)-\DualCost_P(C,h,\beta).
\]

\noindent To prove
\Cref{thm:kmedian}, we use three lemmas.

\noindent First, for a fixed center tuple, it suffices to preserve the dual objective uniformly over all dual
vectors.

\begin{restatable}[Dual preservation implies load preservation]{lemma}{FiniteDualToPrimal} \label{lem:finite-dual-to-primal} Fix \(C\in X^k\). If \[ |\Error_C(\beta)|\le \eta \qquad\text{for every }\beta\in\R^k, \] then \[ |\cost_Q(C,h)-\cost_P(C,h)|\le \eta \qquad\text{for every }h\in n\Delta_k. \] \end{restatable}

Second, for any fixed dual vector, the sampling error satisfies a
standard concentration bound. After fixing \(C\) and \(\beta\), the
coreset contribution is an unbiased estimator of the corresponding
contribution of \(P\). The ring decomposition then provides uniform
bounds on the magnitude of each summand, as well as the aggregate
variance proxy needed for concentration.

\begin{restatable}[Concentration for one dual vector]{lemma}{FiniteKMedianFixedBeta}
\label{lem:finite-kmedian-fixed-beta}
Fix \(C\in X^k\) and \(\beta\in\R^k\). If \(Q\) is produced by
\Cref{alg:finite-kmedian-ring-sampling} with the sample size parameter \(T\)
defined therein, then
\[
|\Error_C(\beta)|\le \eps\opt
\]
with probability at least
\[
1-2\exp\{-\Omega(\eps^2T)\}.
\]
\end{restatable}

Third, the fixed-$\beta$ concentration bound can be made uniform over all
dual vectors. Although $\beta$ ranges over the infinite set
$\mathbb R^k$, the resulting cost vectors have finite combinatorial complexity, so the
uniform guarantee reduces to a union bound over finitely many
concentration events.

\begin{restatable}[Uniform dual preservation for fixed centers]{lemma}{FiniteKMedianUniformBeta}
\label{lem:finite-kmedian-uniform-beta}
Fix \(C\in X^k\). If
\[
T\gtrsim
\frac{k\log N+\log(1/\delta)}{\eps^2},
\]
then, with probability at least \(1-\delta\),
\[
|\Error_C(\beta)|\lesssim  \eps\opt
\qquad\text{for every }\beta\in\R^k .
\]
\end{restatable}

\begin{proof}[Proof of \Cref{thm:kmedian}.]

    Applying \Cref{lem:finite-kmedian-uniform-beta} to a fixed center tuple
\(C\in X^k\), with failure probability \(\delta/N^k\), we obtain a
high-probability event on which the dual error is small uniformly over
all dual vectors \(\beta\in\mathbb R^k\). More precisely, for this fixed
choice of \(C\), the lemma gives
\[
|\Error_C(\beta)| \lesssim  \eps\opt
\qquad
\text{for all } \beta\in\mathbb R^k .
\]
There are at most \(N^k\) possible center tuples \(C\in X^k\).
Therefore, taking a union bound over all such tuples, we get that with
probability at least \(1-\delta\),
\[
|\Error_C(\beta)| \lesssim  \eps\opt
\qquad
\text{for every } C\in X^k
\text{ and every } \beta\in\mathbb R^k .
\]
The sampling parameter required for this event is
\[
T \gtrsim
\frac{k\log N+\log(1/\delta)}{\eps^2}.
\]
Indeed, the only additional cost of making the guarantee simultaneous
over all \(C\in X^k\) is the union bound over at most \(N^k\) choices of
centers, which contributes an additive \(k\log N\) term.

We now condition on the above event. Since the dual error is controlled
for every \(C\in X^k\) and every \(\beta\in\mathbb R^k\),
\Cref{lem:finite-dual-to-primal} implies that the
corresponding  assignment costs are preserved additively.
Thus, for every center tuple \(C\in X^k\) and every  \loadvector
\(h\in n\Delta_k\),
\[
\left|
\operatorname{cost}_Q(C,h)
-
\operatorname{cost}_P(C,h)
\right|
\lesssim  \eps\opt \lesssim \eps \operatorname{cost}_P(C,h). 
 \]
This gives the stated \(\eps\)-coreset
guarantee. 
\end{proof}

The rest of this section proves
\Cref{lem:finite-dual-to-primal},
\Cref{lem:finite-kmedian-fixed-beta}, and
\Cref{lem:finite-kmedian-uniform-beta}.

\subsection{Dual Cost Preservation: Proof of \Cref{lem:finite-dual-to-primal}}

The following lemma shows that preserving the dual objective implies the coreset guarantee.

\FiniteDualToPrimal*

\begin{proof}
By \Cref{lem:finite-strong-duality},
\[
\operatorname{cost}_P(C,h)
=
\sup_{\beta\in\mathbb R^k}
\DualCost_P(C,h,\beta),
\qquad
\operatorname{cost}_Q(C,h)
=
\sup_{\beta\in\mathbb R^k}
\DualCost_Q(C,h,\beta).
\]
For every \(\beta\in\mathbb R^k\),
\[
\DualCost_Q(C,h,\beta)
-
\DualCost_P(C,h,\beta)
=
\Error_C(\beta),
\]
because the term \(-h\cdot\beta\) is identical in the two dual
objectives. By assumption, \(|\Error_C(\beta)|\le \eta\) for every
\(\beta\). Therefore,
\[
\DualCost_Q(C,h,\beta)
\le
\DualCost_P(C,h,\beta)+\eta
\qquad
\forall \beta\in\mathbb R^k .
\]
Taking suprema over \(\beta\) gives
\[
\operatorname{cost}_Q(C,h)
\le
\operatorname{cost}_P(C,h)+\eta .
\]
The same argument with \(P\) and \(Q\) interchanged gives
\[
\operatorname{cost}_P(C,h)
\le
\operatorname{cost}_Q(C,h)+\eta .
\]
Combining the two inequalities yields
\[
\left|
\operatorname{cost}_Q(C,h)
-
\operatorname{cost}_P(C,h)
\right|
\le \eta . 
\]
\end{proof}

The important benefit when analyzing \(\Error_C(\beta)\) is that we no longer consider demand vectors. Once we prove a uniform bound on \(\Error_C(\beta)\)
over all dual vectors \(\beta\), the same event simultaneously preserves
the optimum for every \(h\in n\Delta_k\) and no union bound over \loadvectors is needed.


\subsection{Concentration for a fixed dual vector: Proof of \Cref{lem:finite-kmedian-fixed-beta}}

\FiniteKMedianFixedBeta*

\begin{proof}
The function \(f_{C,\beta}\) is the minimum of \(1\)-Lipschitz functions,
so it is \(1\)-Lipschitz.  Thus, for every ring \(R\in\mathcal R\) and
every point \(p\in R\),
\[
|f_{C,\beta}(p)-f_{C,\beta}(a_R)|\le d(p,a_R)\le \rho_R .
\]

Fix a ring \(R\).  Let \(q_{R,1},\ldots,q_{R,m_R}\) be
the independent uniform samples from \(R\).  Set
\[
g_R(p)=f_{C,\beta}(p)-f_{C,\beta}(a_R),
\qquad
\bar g_R=\frac1{n_R}\sum_{p\in R}g_R(p).
\]
The contribution of \(R\) to the sampling error is
\[
\sum_{t=1}^{m_R}\frac{n_R}{m_R}\bigl(g_R(q_{R,t})-\bar g_R\bigr).
\]
The summands are independent and mean zero.  Each has magnitude at most
\(2n_R\rho_R/m_R\).  Hoeffding's inequality (see \ref{lem:hoeffding}) gives, for every \(s>0\),
\[
\Pr\{|\operatorname{Err}_C(\beta)|\ge s\}
\le
2\exp\left(
-\Omega\left(
\frac{s^2}{\sum_{R\in\mathcal R}(n_R\rho_R)^2/m_R}
\right)
\right).
\]
The sampling rule implies
\[
\sum_{R\in\mathcal R}\frac{(n_R\rho_R)^2}{m_R}
\le 
\frac{\Wone^2}{T}.
\]
By \Cref{lem:finite-kmedian-ring-scale}, \(\Wone\lesssim\opt\).  Taking
\(s=\eps\opt\) proves the lemma.
\end{proof}


\subsection{Uniform Preservation over Dual Vectors: Proof of \Cref{lem:finite-kmedian-uniform-beta}}

It remains to make the fixed-\(\beta\) concentration statement uniform
over all dual vectors. Fix a center tuple
\(C\in X^k\). Our goal in this subsection is to prove that, with high
probability,
\[
|\Error_C(\beta)| \lesssim \varepsilon \OPT
\qquad \text{for every } \beta\in\mathbb R^k .
\]
The difficulty is that \(\beta\) ranges over an infinite set. We overcome
this by showing that the sampling error depends on \(\beta\) only through
a residual vector, and that the set of all such residual vectors admits a small discretization.

\noindent
For a ring \(R\) and a point \(p\in R\), define the normalized residual
\[
G_{C,\beta}(R,p)
=
\frac{f_{C,\beta}(p)-f_{C,\beta}(a_R)}{\rho_R}.
\]
Since \(f_{C,\beta}\) is \(1\)-Lipschitz and \(d(p,a_R)\le \rho_R\), we
have
\(
|G_{C,\beta}(R,p)|\le 1.
\)
Thus,
\[
G_{C,\beta}\in[-1,1]^{\mathcal D},
\qquad
\mathcal D=\{(R,p):R\in\mathcal R,\ p\in R\}.
\]

The following observation provides the deterministic link between
approximation in the residual space and approximation of the dual sampling
error. First, after centering at the anchor of
each ring, the anchor terms cancel exactly. Indeed, the sampled total weight
from \(R\) is exactly \(n_R\), so the contribution of the constant value \(f_{C,\beta}(a_R)\) is the same
from the sampled copy of \(R\) in \(Q\) and from the original ring \(R\). Therefore
\[
\Error_C(\beta)
=
\sum_{R\in\mathcal R}\rho_R
\left(
\frac{n_R}{m_R}\sum_{t=1}^{m_R}G_{C,\beta}(R,q_{R,t})
-
\sum_{p\in R}G_{C,\beta}(R,p)
\right).
\]
Thus \(\Error_C(\beta)\) is controlled entirely by the normalized residual
vector \(G_{C,\beta}\).

Second, this dependence is Lipschitz in the \(\ell_\infty\) norm. Suppose
\(\beta,\beta'\in\mathbb R^k\) satisfy
\[
\|G_{C,\beta}-G_{C,\beta'}\|_\infty\le \gamma .
\]
Then, for every realization of the sample,

\begin{equation}
 \label{eq:error-due-to-net}
\begin{aligned}
|\Error_C(\beta)-\Error_C(\beta')|
&\le
\sum_{R\in\mathcal R}\rho_R
\left(
\frac{n_R}{m_R}\sum_{t=1}^{m_R}\gamma
+
\sum_{p\in R}\gamma
\right) \\
&=
2\gamma\sum_{R\in\mathcal R} n_R\rho_R
=
2\gamma W_1 .
\end{aligned}
\end{equation}

Consequently, if every \(G_{C,\beta}\) is within
\(\gamma\) in \(\ell_\infty\) distance of some representative
\(G_{C,\widehat\beta}\), then concentration for the representatives extends
to all dual vectors with an additional loss of at most \(2\gamma W_1\).
Since \(W_1\lesssim \OPT\), choosing \(\gamma\) to be a sufficiently small
constant multiple of \(\varepsilon\) makes this loss \(O(\varepsilon)\OPT\).
It remains to show that the residual vectors induced by dual solutions admit
a small such net.

\begin{lemma}[Net for residual vectors]
\label{lem:kmedian-residual-net}
For every \(\gamma\in(0,1)\), there is a set
\(\mathcal B_C(\gamma)\subseteq\mathbb R^k\) with
\[
|\mathcal B_C(\gamma)|
\le
\left(\frac{Nk}{\gamma}\right)^{O(k)}
\]
such that for every \(\beta\in\mathbb R^k\), there exists
\(\widehat\beta\in\mathcal B_C(\gamma)\) satisfying
\[
\|G_{C,\beta}-G_{C,\widehat\beta}\|_\infty\le \gamma .
\]
\end{lemma}

\begin{proof}
We first bound the number of possible minimizer patterns. Since every input
point and every anchor lie in \(X\), it suffices to consider the values of
\(f_{C,\beta}\) on points of \(X\). For every \(z\in X\) and every pair
\(i,j\in[k]\), consider the hyperplane
\[
d(z,c_i)+\beta_i=d(z,c_j)+\beta_j
\]
in \(\mathbb R^k\). There are \(M=O(Nk^2)\) such hyperplanes.

The induced arrangement decomposes \(\mathbb R^k\) into faces on which all
comparisons
\[
d(z,c_i)+\beta_i
\quad\text{versus}\quad
d(z,c_j)+\beta_j
\]
are fixed, allowing equality. After applying a fixed tie-breaking rule, each
face therefore determines a single minimizer pattern
\[
\pi_\beta(z)
\in
\arg\min_{i\in[k]}\{d(z,c_i)+\beta_i\},
\qquad z\in X.
\]
The number of faces in an arrangement of \(M\) hyperplanes in
\(\mathbb R^k\) is \(M^{O(k)}\). Hence the number of possible minimizer
patterns is at most
\[
(O(Nk^2))^{O(k)}
=
\exp\{O(k\log(Nk))\}.
\]

\noindent
Fix one such pattern \(\pi\). For every \(\beta\) inducing \(\pi\),
\[
f_{C,\beta}(z)=d(z,c_{\pi(z)})+\beta_{\pi(z)}
\qquad
\forall z\in X.
\]
Thus, for every \((R,p)\in\mathcal D\),
\[
G_{C,\beta}(R,p)
=
\frac{
d(p,c_{\pi(p)})-d(a_R,c_{\pi(a_R)})
+
\beta_{\pi(p)}-\beta_{\pi(a_R)}
}{\rho_R}.
\]
Therefore, once \(\pi\) is fixed, the residual vector \(G_{C,\beta}\) is an
affine function of \(\beta\). The set of residual vectors inducing this
pattern lies in an affine subspace of dimension at most \(k\), equipped with
the norm inherited from \(\ell_\infty(\mathcal D)\), and also lies in the
cube \([-1,1]^{\mathcal D}\).

By the standard volumetric covering argument
\cite[Proposition~4.2.10]{vershynin2026high}, applied inside the
\(k\)-dimensional affine hull with the induced \(\ell_\infty\) norm, any subset
of a \(k\)-dimensional affine subspace with \(\ell_\infty\)-diameter at most
\(2\) has a \(\gamma\)-net of size at most
\[
\left(\frac{C_0}{\gamma}\right)^k
\]
where \(C_0>0\) is a universal constant. Applying this bound to the residual
vectors inducing the fixed pattern gives such a net. Moreover, by taking a
maximal \(\gamma\)-separated subset of the induced residual vectors, the net
points may be chosen to be induced by actual dual vectors.

Taking the union over all minimizer patterns gives a net of size at most
\[
\exp\{O(k\log(Nk))\}\cdot
\left(\frac{C_0}{\gamma}\right)^k
\le
\left(\frac{Nk}{\gamma}\right)^{O(k)} .
\]
Choosing one dual vector representative for each net point gives
\(\mathcal B_C(\gamma)\).
\end{proof}

We now use the residual net to generalize the fixed-\(\beta\) concentration
bound to a statement that holds simultaneously for every dual vector.

\FiniteKMedianUniformBeta*

\begin{proof}
Let \(\gamma=c_0\eps\), where \(c_0\) is a sufficiently small constant.
Apply \Cref{lem:finite-kmedian-fixed-beta} to every
\(\widehat\beta\in\mathcal B_C(\gamma)\), and take a union bound using
\Cref{lem:kmedian-residual-net}. With probability at least \(1-\delta\),
\[
|\Error_C(\widehat\beta)|\le \frac12\eps\opt
\qquad\text{for every }\widehat\beta\in\mathcal B_C(\gamma),
\]
provided that
\[
T\gtrsim
\frac{k\log N +\log(1/\delta)}{\eps^2}.
\]

Now fix an arbitrary \(\beta\in\mathbb R^k\), and choose
\(\widehat\beta\in\mathcal B_C(\gamma)\) with
\[
\|G_{C,\beta}-G_{C,\widehat\beta}\|_\infty\le \gamma .
\]
By \eqref{eq:error-due-to-net}, we have
\[
|\Error_C(\beta)-\Error_C(\widehat\beta)|
\le
2\gamma\Wone .
\]
\noindent
Using \(\Wone\lesssim\opt\), and choosing \(c_0\) sufficiently small, this is
at most \(\eps\opt/2\). Therefore,
\[
|\Error_C(\beta)|
\le
|\Error_C(\widehat\beta)|
+
|\Error_C(\beta)-\Error_C(\widehat\beta)|
\le
\eps\opt .
\]
Since \(\beta\) was arbitrary, the bound holds uniformly over
\(\mathbb R^k\).
\end{proof}

\section{Finite-Metric Capacitated $\kmeans$}

We next prove the coreset theorem for capacitated
$\kmeans$ in finite metrics. 

\begin{theorem}[Finite-metric fractional capacitated $\kmeans$]
\label{thm:kmeans}
For every $\eps,\delta\in(0,1)$, there is a randomized algorithm which
outputs a weighted subset $Q\subseteq P$ of size
\[
|Q|
=
O\left(
\frac{k\log N+\log(1/\delta)}{\eps^2}
\right)
\]
such that, with probability at least $1-\delta$, $Q$ is an
$\eps$-coreset for finite-metric capacitated $\kmeans$.
\end{theorem}

\paragraph{Coreset Construction.}
The construction is the same as in the $\kmedian$ section, with squared
distances replacing distances.

Let $A=\{a_1,\ldots,a_k\}$ be a constant-factor approximate
unconstrained $\kmeans$ solution for $P$. The points of $A$ are called
anchors. Assign each point $p\in P$ to its nearest anchor, breaking ties
arbitrarily. For an anchor $a_j$, let $P_j$ be its assigned cluster, and set
$\Phi_j=\sum_{p\in P_j}d(p,a_j)^2$. Then
$\sum_{j\in [k]}\Phi_j \lesssim \opt$, where $\opt$ is the optimum
unconstrained $\kmeans$ cost of $P$.

We assume without loss of generality that no cluster has zero reference
cost. Indeed, if \(\Phi_j=0\), then all points in the cluster coincide with the
anchor, so we store a single point with the appropriate weight. This
represents the cluster exactly and increases the coreset size by at most \(k\).

\paragraph{Ring Partitioning.}
For each anchor \(a_j\), set \(
\lambda_j=\frac{\Phi_j}{|P_j|}.
\)
We partition \(P_j\) by squared distance from its anchor. Define the
innermost ring by
\[
R_{j,0}
=
\{p\in P_j : d(p,a_j)^2\le \lambda_j\},
\]
and for \(t\ge 1\), define
\[
R_{j,t}
=
\{p\in P_j : 2^{t-1}\lambda_j<d(p,a_j)^2\le 2^t\lambda_j\}.
\]
Empty rings are discarded.

\noindent
Let \(\mathcal R\) denote the collection of all nonempty rings. We call these the sampled rings.
For a sampled ring \(R=R_{j,t}\), write
\[
a_R=a_j,
\qquad
n_R=|R|,
\qquad
\rho_R^2=2^t\lambda_j.
\]
Thus,
\[
d(p,a_R)^2\le \rho_R^2
\qquad
\forall p\in R.
\]

The quantity \(n_R\rho_R^2\) is an upper bound, up to constants, on the total
anchor cost scale of ring \(R\). We use these quantities as the sampling
priorities of the rings. We normalize these scales by setting
\[
\Wtwo
=
\sum_{R\in\mathcal R} n_R\rho_R^2 .
\]
The algorithm will allocate samples to each ring \(R\) proportionally to
\(n_R\rho_R^2/\Wtwo\).

Each anchor contributes at most \(O(\log n)\) sampled rings. Indeed, if
\(R_{j,t}\) is nonempty for some \(t\ge 1\), then some point in the ring
has squared distance larger than \(2^{t-1}\lambda_j\) from \(a_j\). Since
the total squared cost of the cluster is \(\Phi_j=|P_j|\lambda_j\), we have
\[
2^{t-1}\lambda_j < \Phi_j=|P_j|\lambda_j,
\]
and hence \(t=O(\log |P_j|)\). Therefore,
\[
|\mathcal R|=O(k\log n).
\]

\paragraph{Sampling.}
For each  ring $R$, sample
$m_R=\left\lceil T n_R\rho_R^2/\Wtwo\right\rceil$ points independently and
uniformly from $R$, and give each sampled point weight $n_R/m_R$. Here
$T=\Theta((k\log N+\log(1/\delta))/\eps^2)$. The total number of points sampled in the coreset is $O(T)$. 

\begin{claim}
\label{lem:finite-kmeans-ring-scale}
The squared ring scales satisfy
\(
\Wtwo=\sum_{R\in\mathcal R}n_R\rho_R^2\lesssim \opt .
\)
\end{claim}
\noindent The proof of the above claim is identical to \Cref{lem:finite-kmedian-ring-scale}
with squared distances replacing distances.


\subsection{Analysis}

The proof follows the same duality template as the \(\kmedian\) analysis,
but squared distances introduce an additional difficulty. After centering
at a ring anchor, the difference
\[
d(p,c)^2-d(a_R,c)^2
\]
is not controlled only by the squared ring scale \(\rho_R^2\); it also
contains a cross term depending on the distance from the anchor \(a_R\) to
the candidate center \(c\). This prevents a direct  application of  Hoeffding's inequality to ensure concentration.

First, we introduce a truncated ring-labelled objective. For centers that
are far from a ring, we replace the point-center squared distance by the
corresponding anchor-center squared distance. This changes every feasible
assignment cost by only a \(1\pm O(\eps)\) factor, while restoring bounded variation
within a ring. This is in fact enough to show that coresets of size \(\widetilde O(k\eps^{-4})\) suffice.

Second, we use a weak dual convergence argument. Unlike in the
\(\kmedian\) proof, we do not approximate every dual vector to the same
additive accuracy. Dual vectors that are close to optimal are preserved
accurately while dual vectors far from optimal may be approximated more
coarsely, because their dual value is already too small to
affect the maximum. This is the step that recovers the
\(\widetilde O(k\eps^{-2})\) sample dependence.

We now make these two ingredients formal. The proof is organized around three lemmas.
First, we define the truncated objective formally and show that it is a good approximation for the original \(\kmeans\) objective.

\paragraph{Truncation.}
The truncation step separates the case where a candidate center is much
farther from the anchor than the scale of the ring. In this case, replacing
the point-center cost by the corresponding anchor-center cost changes the
cost only by a \(1\pm O(\eps)\) factor. This restores bounded within-ring
variation for the truncated objective.

\begin{definition}[Truncated costs]
\label{def:kmeans-truncated-costs}
For a point \(p\), define the truncated cost of assigning \(p\) to center
\(c_i\) as follows. If \(p\) belongs to a sampled ring \(R\in\mathcal R\),
set
\[
\widetilde \ell_i(p)
=
\begin{cases}
d(p,c_i)^2, & \text{if } d(a_R,c_i)\le \rho_R/\eps,\\
d(a_R,c_i)^2, & \text{if } d(a_R,c_i)> \rho_R/\eps .
\end{cases}
\]
If \(p\) belongs to a cluster that is represented exactly, set
\(
\widetilde\ell_i(p)=d(p,c_i)^2 .
\)
When evaluating the centered anchor term for a sampled ring \(R\), we set
\(
\widetilde\ell_i(a_R):=d(a_R,c_i)^2 .
\)

For a weighted instance \(S\) supported on \(P\) and a \loadvector
\(h\in n\Delta_k\), let \(\TCost_S(C,h)\) denote the optimum value of the
fractional assignment LP on \(S\) with \loadvector \(h\), where assigning
\(p\) to \(c_i\) has cost \(\widetilde\ell_i(p)\).
\end{definition}

\begin{restatable}[Truncation comparison]{lemma}{KMeansTruncationComparison}
\label{lem:kmeans-truncation-comparison}
For every weighted ring-labelled instance \(S\), every center tuple
\(C\in X^k\), and every \loadvector \(h\in n\Delta_k\),
\[
\TCost_S(C,h)=(1\pm O(\eps))\cost_S(C,h).
\]
\end{restatable}

\noindent
We next formulate the weak dual convergence condition. For the rest of
the \(\kmeans\) analysis, all dual quantities are defined with respect to
the truncated costs \(\widetilde \ell_i\).

\paragraph{Dual formulation.}
We use the same dual formulation as in the \(\kmedian\) section, with the original costs being replaced by the truncated costs \(\widetilde\ell_i\). Thus the truncated objective \(\TCost_S(C,h)\) can be written as the supremum of a reduced dual objective over \(\beta\in\mathbb R^k\). As before, when comparing \(P\) and \(Q\), the load term \(-h\cdot\beta\) cancels. Therefore the sampling error depends only on \(C\) and \(\beta\), not on the \loadvector \(h\).

\paragraph{Weak dual convergence.}
The weak dual convergence argument uses a nonuniform notion of dual
preservation. A dual vector that is nearly optimal for the
instance must be preserved accurately, while a dual vector whose population
dual value is far below the optimum may be preserved more coarsely.

\begin{definition}[Dual gap and allowed error]
\label{def:kmeans-gap-allowed-error}
Fix \(C\in X^k\), \(h\in n\Delta_k\), and \(\beta\in\mathbb R^k\). The
truncated dual gap of \(\beta\) with respect to \(C,h\) is
\[
\Delta_{C,h}(\beta)
=
\TCost_P(C,h)-\DualCost_P(C,h,\beta).
\]
By strong duality, \(\Delta_{C,h}(\beta)\ge 0\).

\noindent
For fixed \(C\) and \(\beta\), define the allowed error
\[
\Allow(\beta)
=
\inf_{h\in n\Delta_k}
\left\{
\eps\,\TCost_P(C,h)
+
\frac14\Delta_{C,h}(\beta)
\right\}.
\]
\end{definition}
\noindent
The next lemma explains why this is the right notion of dual control. It
says that controlling the sampling error only up to the allowed error is
already enough to preserve the truncated exact-load objective.

\begin{restatable}[Allowed-error dual preservation]{lemma}{KMeansAllowedErrorPreservation}
\label{lem:kmeans-allowed-error-preservation}
Fix \(C\in X^k\). Suppose that
\[
|\Error_C(\beta)|\le \Allow(\beta)
\qquad
\text{for every }\beta\in\mathbb R^k .
\]
Then, for every \loadvector \(h\in n\Delta_k\),
\[
\TCost_Q(C,h)=(1\pm O(\eps))\TCost_P(C,h).
\]
\end{restatable}

It remains to prove that the sampled instance satisfies the allowed-error
condition. This is the main sampling statement for \(\kmeans\). It is the
step that improves the sample dependence from the naive
\(\widetilde O(k\eps^{-4})\) bound obtained from uniform Hoeffding's inequality
to the desired \(\widetilde O(k\eps^{-2})\) bound.


\begin{restatable}[Uniform weak dual convergence for fixed centers]{lemma}{KMeansWeakDualConvergence}
\label{lem:kmeans-weak-dual-convergence}
Fix \(C\in X^k\). If
\[
T\gtrsim
\frac{k\log N+\log(1/\delta)}{\eps^2},
\]
then, with probability at least \(1-\delta\),
\[
|\Error_C(\beta)|\le \Allow(\beta)
\qquad
\text{for every }\beta\in\mathbb R^k.
\]
\end{restatable}

\begin{proof}[Proof of \Cref{thm:kmeans}.]
Apply \Cref{lem:kmeans-weak-dual-convergence} to each fixed
\(C\in X^k\) with failure probability \(\delta/N^k\).  Since there are at
most \(N^k\) center tuples, a union bound gives, with probability at least
\(1-\delta\),
\[
|\Error_C(\beta)|\le \Allow(\beta)
\qquad
\text{for every } C\in X^k
\text{ and every } \beta\in\mathbb R^k .
\]
On this event, \Cref{lem:kmeans-allowed-error-preservation} gives
\[
\TCost_Q(C,h)
=
(1\pm O(\eps))\TCost_P(C,h)
\qquad
\text{for every }C\in X^k
\text{ and every }h\in n\Delta_k .
\]

It remains to pass from truncated costs to the original costs.  By
\Cref{lem:kmeans-truncation-comparison}, applied to both \(P\) and \(Q\),
\[
\TCost_P(C,h)=(1\pm O(\eps))\cost_P(C,h),
\qquad
\TCost_Q(C,h)=(1\pm O(\eps))\cost_Q(C,h).
\]
Combining these, we have
\[
\cost_Q(C,h)
=
(1\pm O(\eps))\cost_P(C,h)
\]
simultaneously for every \(C\in X^k\) and every \(h\in n\Delta_k\).
Rescaling \(\eps\) by a sufficiently small constant gives the claimed
\(\eps\)-coreset guarantee.
\end{proof}

The rest of the section proves
Lemma~\ref{lem:kmeans-truncation-comparison},
Lemma~\ref{lem:kmeans-allowed-error-preservation}, and
Lemma~\ref{lem:kmeans-weak-dual-convergence}.

\subsection{Truncated Costs Approximate True Costs: Proof of Lemma~\ref{lem:kmeans-truncation-comparison}}

We now show that the truncated objective is equivalent to the original
squared-distance objective. The reason is simple: if a point-center cost is truncated, the
center is far enough from the ring that every point in the ring has
nearly the same squared distance to it as the anchor does.

\KMeansTruncationComparison*

\begin{proof}
Fix a weighted instance $S$ supported on $P$, a center tuple $C$, and a
\loadvector $h$. We first show that, for every point $p$ and
center $c_i$,
\begin{equation}
\label{eq:kmeans-truncation-pointwise}
\widetilde\ell_i(p)=(1\pm O(\eps))d(p,c_i)^2 .
\end{equation}

If the cost of assigning $p$ to $c_i$ is not truncated, then
$\widetilde\ell_i(p)=d(p,c_i)^2$ and there is nothing to prove. If it is
truncated, then $\widetilde\ell_i(p)=d(a_R,c_i)^2$ and
$d(a_R,c_i)>\rho_R/\eps$. Since $d(p,a_R)\le \rho_R$, we have
$d(p,a_R)\le \eps d(a_R,c_i)$. Hence, by the triangle inequality,
$d(p,c_i)=(1\pm O(\eps))d(a_R,c_i)$, and therefore
\[
d(p,c_i)^2=(1\pm O(\eps))d(a_R,c_i)^2
=
(1\pm O(\eps))\widetilde\ell_i(p).
\]
This proves \eqref{eq:kmeans-truncation-pointwise} after adjusting the
implicit constant.

Now take any feasible fractional assignment $x$ for $S$ with load
vector $h$. Since $x_{p,i}\ge 0$, summing
\eqref{eq:kmeans-truncation-pointwise} over all assigned pairs $(p,i)$ gives
\[
\sum_{p,i}x_{p,i}\widetilde\ell_i(p)
=
(1\pm O(\eps))
\sum_{p,i}x_{p,i}d(p,c_i)^2 .
\]
Applying this once to an optimal assignment for
$\operatorname{cost}_S(C,h)$ and once to an optimal assignment for
$\TCost_S(C,h)$ gives
\[
\TCost_S(C,h)
=
(1\pm O(\eps))\operatorname{cost}_S(C,h).
\]
\end{proof}

 \subsection{From Dual Error to Cost Preservation: Proof of \Cref{lem:kmeans-allowed-error-preservation}} We prove that controlling the dual sampling error by the allowed error \(\Allow(\beta)\) is enough to preserve the truncated assignment cost for every demand vector. \KMeansAllowedErrorPreservation*

\begin{proof} 
Fix \(h\in n\Delta_k\). 
For the upper bound, for every \(\beta\in\mathbb R^k\), \[ \DualCost_Q(C,h,\beta) \le \DualCost_P(C,h,\beta)+\Allow(\beta). \]

By the definition of \(\Allow(\beta)\), \[ \Allow(\beta) \le \eps\,\TCost_P(C,h) + \frac14\Delta_{C,h}(\beta). \]

Therefore, 
\[ 
\begin{aligned}
\DualCost_Q(C,h,\beta) &\le \DualCost_P(C,h,\beta) + \eps\,\TCost_P(C,h) + \frac14\bigl(\TCost_P(C,h)-\DualCost_P(C,h,\beta)\bigr) \\ &= \frac34\DualCost_P(C,h,\beta) + \left(\frac14+\eps\right)\TCost_P(C,h) \\ &\le (1+\eps)\TCost_P(C,h), 
\end{aligned} 
\] where the last inequality uses \(\DualCost_P(C,h,\beta)\le \TCost_P(C,h)\).

Taking the supremum over \(\beta\) gives \[ \TCost_Q(C,h)\le (1+\eps)\TCost_P(C,h). \] \noindent For the lower bound, let \(\beta^\star\) be an optimal dual solution for \(\TCost_P(C,h)\). Then \(\Delta_{C,h}(\beta^\star)=0\), and hence \[ \Allow(\beta^\star)\le \eps\,\TCost_P(C,h). \] Thus \[ \TCost_Q(C,h) \ge \DualCost_Q(C,h,\beta^\star) \ge \DualCost_P(C,h,\beta^\star)-\Allow(\beta^\star) \ge (1-\eps)\TCost_P(C,h). \] 
\end{proof}

\subsection{Weak Dual Convergence:
Proof of Lemma~\ref{lem:kmeans-weak-dual-convergence}}

The proof of weak dual convergence is the technical core of the
\(\kmeans\) analysis. We prove the statement for a fixed center tuple
\(C\), uniformly over all dual vectors \(\beta\). The extension to all
\(C\in X^k\) is then obtained by a union bound in the proof of
\Cref{thm:kmeans}.

The proof has two internal ingredients. First, a localized concentration
bound shows that, for one fixed dual vector, the sampling error is
controlled at the allowed-error scale. Second, a discretization argument
reduces the infinite family of dual vectors to a finite representative
set. The point of the localized argument is that near-optimal dual
vectors are controlled accurately, while dual vectors with large dual gap
may be controlled more coarsely.

The first ingredient gives the desired allowed-error bound for one fixed
dual vector.

\begin{lemma}[Pointwise concentration]
\label{lem:kmeans-pointwise-concentration}
Fix \(C\in X^k\), \(\beta\in\mathbb R^k\), and
\(\delta\in(0,1)\). If
\[
T
\gtrsim
\frac{\log(1/\delta)}{\eps^2},
\]
then
\[
|\Error_C(\beta)|
\le
\frac18 \Allow(\beta)
\]
with probability at least \(1-\delta\).
\end{lemma}

The second ingredient upgrades control on finitely many representative
dual vectors to control over all \(\beta\in\mathbb R^k\).

\begin{lemma}[Dual-vector discretization]
\label{lem:kmeans-dual-vector-discretization}
Fix \(C\in X^k\). There is a set
\(\mathcal B_C\subseteq\mathbb R^k\) of size
\[
|\mathcal B_C|
\le
\exp\{O(k\log(N/\eps))\}
\]
such that the following implication holds. If
\[
|\Error_C(\beta)|\le \frac18 \Allow(\beta)
\qquad
\forall \beta\in\mathcal B_C,
\]
then
\[
|\Error_C(\beta)|\le \Allow(\beta)
\qquad
\forall \beta\in\mathbb R^k .
\]
\end{lemma}

With the two lemmas in hand, we can state and prove the uniform convergence for all dual vectors. We return to the proofs of
the lemmas after that.

\KMeansWeakDualConvergence*
\begin{proof}
Fix \(C\in X^k\), and let \(\mathcal B_C\) be the representative set from
\Cref{lem:kmeans-dual-vector-discretization}. Apply
\Cref{lem:kmeans-pointwise-concentration} to every
\(\beta\in\mathcal B_C\), using failure probability
\[
\delta'
=
\frac{\delta}{|\mathcal B_C|}.
\]
By a union bound over \(\mathcal B_C\), with probability at least
\(1-\delta\),
\[
|\Error_C(\beta)|
\le
\frac18 \Allow(\beta)
\qquad
\forall \beta\in\mathcal B_C,
\]
provided
\[
T
\gtrsim
\frac{k\log N+\log(1/\delta)}{\eps^2}.
\]
On this event, \Cref{lem:kmeans-dual-vector-discretization} gives
\[
|\Error_C(\beta)|
\le
\Allow(\beta)
\qquad
\forall \beta\in\mathbb R^k .
\]
This proves \Cref{lem:kmeans-weak-dual-convergence}.
\end{proof}

\subsubsection{Pointwise concentration}

We prove \Cref{lem:kmeans-pointwise-concentration}. Throughout this
subsection, fix a center tuple \(C\in X^k\) and a dual vector
\(\beta\in\mathbb R^k\). 

For a ring \(R\), define
\[
g_R(p)=f_{C,\beta}(p)-f_{C,\beta}(a_R),
\qquad
\bar g_R=\frac1{n_R}\sum_{p\in R}g_R(p).
\]
The truncation step gives the centered range bound
\begin{equation}
\label{eq:kmeans-centered-range}
|g_R(p)|
\lesssim \eps^{-1} \rho_R^2
\qquad
\text{for every }p\in R .
\end{equation}

We also use the following variance-gap estimate, proved in the next
subsection.

\begin{lemma}[Variance-gap estimate]
\label{lem:kmeans-variance-gap-estimate}
Fix \(C\in X^k\), \(h\in n\Delta_k\), and
\(\beta\in\mathbb R^k\). Then
\[
\sum_{R\in\mathcal R}
\frac{n_R}{\rho_R^2}
\operatorname{Var}_{p\sim R}[g_R(p)]
\le
O\!\left(
\TCost_P(C,h)
+
\eps^{-1}\Delta_{C,h}(\beta)
\right).
\]
\end{lemma}

This estimate says that large within-ring variance can occur only for
dual vectors with large dual gap. Such vectors have a larger allowed
error in \(\Allow(\beta)\), which is exactly what Bernstein concentration
exploits.

\begin{proof}[Proof of \Cref{lem:kmeans-pointwise-concentration}]
Let \(h^\star\in n\Delta_k\) attain the infimum in the definition of
\(\Allow(\beta)\), and write
\[
U=\TCost_P(C,h^\star),
\qquad
D=\Delta_{C,h^\star}(\beta).
\]
Thus
\[
\Allow(\beta)=\eps U+\frac14D.
\]

For each ring \(R\), let
\(q_{R,1},\ldots,q_{R,m_R}\) be the independent uniform samples from
\(R\). Since the sampled points from \(R\) have total weight exactly
\(n_R\), the anchor contribution cancels. Hence
\begin{equation}
\label{eq:kmeans-error-centered-ring}
\Error_C(\beta)
=
\sum_{R\in\mathcal R}
n_R
\left(
\frac1{m_R}\sum_{t=1}^{m_R}g_R(q_{R,t})
-
\bar g_R
\right).
\end{equation}
Equivalently,
\[
\Error_C(\beta)
=
\sum_{R\in\mathcal R}\sum_{t=1}^{m_R} Z_{R,t},
\qquad
Z_{R,t}
=
\frac{n_R}{m_R}
\bigl(g_R(q_{R,t})-\bar g_R\bigr).
\]
The variables \(Z_{R,t}\) are independent and mean zero.

We now record the two quantities needed for Bernstein's inequality. First,
using the sampling rule
\[
m_R\asymp T\frac{n_R\rho_R^2}{W_2},
\]
we have
\[
\begin{aligned}
\sum_{R,t}\mathbb E[Z_{R,t}^2]
&=
\sum_{R\in\mathcal R}
\frac{n_R^2}{m_R}
\operatorname{Var}_{p\sim R}[g_R(p)]  \\
&\le
O\!\left(
\frac{W_2}{T}
\sum_{R\in\mathcal R}
\frac{n_R}{\rho_R^2}
\operatorname{Var}_{p\sim R}[g_R(p)]
\right).
\end{aligned}
\]
By \Cref{lem:kmeans-variance-gap-estimate}, applied with
\(h=h^\star\), and since \(W_2\lesssim U\),
\begin{equation}
\label{eq:kmeans-bernstein-variance}
V
:=
\sum_{R,t}\mathbb E[Z_{R,t}^2]
\le
O\!\left(
\frac{U}{T}\left(U+\eps^{-1}D\right)
\right).
\end{equation}

Second, by \eqref{eq:kmeans-centered-range},
\[
|g_R(q_{R,t})-\bar g_R|
\lesssim
\eps^{-1} \rho_R^2 .
\]
Using the sampling rule again and \(W_2\lesssim U\), we get
\begin{equation}
\label{eq:kmeans-bernstein-range}
M
:=
\max_{R,t}|Z_{R,t}|
\le
O\!\left(
\frac{\eps^{-1}U}{T}
\right).
\end{equation}

\noindent
Using Bernstein's inequality (see \Cref{lem:bernstein}) applied to the
independent mean-zero variables \(Z_{R,t}\),
\[
|\Error_C(\beta)|
\lesssim
\sqrt{V\log(1/\delta)}
+
M\log(1/\delta)
\]
with probability at least \(1-\delta\), where
\[
V=\sum_{R,t}\mathbb E[Z_{R,t}^2],
\qquad
M=\max_{R,t}|Z_{R,t}|.
\]
Substituting \eqref{eq:kmeans-bernstein-variance} and
\eqref{eq:kmeans-bernstein-range}, we obtain
\[
|\Error_C(\beta)|
\lesssim
\sqrt{
\frac{U}{T}
\left(
U+\eps^{-1}D
\right)\log(1/\delta)
}
+
\frac{\eps^{-1}U}{T}\log(1/\delta).
\]

If
\(
T\gtrsim \frac{\log(1/\delta)}{\eps^2},
\)
then the second term is at most \(O(\eps U)\), with an arbitrarily small
implicit constant after increasing the constant in the lower bound on
\(T\). The same assumption gives
\[
\sqrt{
\frac{U}{T}
\left(
U+\eps^{-1}D
\right)\log(1/\delta)
}
\lesssim
\eps\sqrt{U\left(U+\eps^{-1}D\right)}
\lesssim
\eps U+D,
\]
again with an arbitrarily small implicit constant after increasing the
constant in the lower bound on \(T\). Therefore,
\[
|\Error_C(\beta)|
\le
\frac18\left(\eps U+\frac14D\right)
=
\frac18 \Allow(\beta).
\]
This proves the claim.
\end{proof}

\subsubsection{Proof of the variance-gap estimate}


We now prove the variance bound in \Cref{lem:kmeans-variance-gap-estimate}. This lemma provides the variance term needed for Bernstein's inequality:
although the worst-case within-ring variation of the truncated dual function
can be as large as \(O(\eps^{-1}\rho_R^2)\), its aggregate variance is much
smaller, namely controlled by
\[
\TCost_P(C,h)+\eps^{-1}\Delta_{C,h}(\beta).
\]
Consequently, when the duality gap is small, Bernstein concentration occurs
at the scale of the clustering cost rather than at the scale suggested by
the worst-case range bound alone.

Fix \(C,h,\beta\).  The function
\[
f_{C,\beta}(p)=\min_i\{\widetilde\ell_i(p)+\beta_i\}
\]
chooses, for each point \(p\), the best \(\beta\)-shifted center.  This
minimizing center may change from point to point, so we compare it to the
optimal primal assignment for the fixed \loadvector \(h\).  Let \(x^\star\)
be an optimal assignment attaining \(\TCost_P(C,h)\).

\noindent
For \(p\in P\) and \(i\in[k]\), define the slack
\begin{equation}
\label{eq:kmeans-slack}
s_{p,i}
=
\widetilde\ell_i(p)+\beta_i-f_{C,\beta}(p).
\end{equation}
Thus \(s_{p,i}\) measures how far center \(i\) is from being a minimizer in
the definition of \(f_{C,\beta}(p)\).  Since
\(
f_{C,\beta}(p)=\min_i\{\widetilde\ell_i(p)+\beta_i\},
\)
we have \(s_{p,i}\ge 0\) for every \(p\) and \(i\).

\noindent
The point of this definition is that the duality gap is exactly the total
slack incurred by the optimal primal assignment \(x^\star\):
\begin{equation}
\label{eq:kmeans-gap-equals-slack}
\Delta_{C,h}(\beta)
=
\sum_{p\in P}\sum_{i=1}^k x^\star_{p,i}s_{p,i}.
\end{equation}
Indeed, using the optimality of \(x^\star\) for \(\TCost_P(C,h)\) and the
definition of the dual objective,
\[
\begin{aligned}
\Delta_{C,h}(\beta)
&=
\TCost_P(C,h)-\DualCost_P(C,h,\beta) \\
&=
\sum_{p,i}x^\star_{p,i}\widetilde\ell_i(p)
-
\sum_{p\in P}f_{C,\beta}(p)
+
\sum_{i=1}^k h_i\beta_i \\
&= \sum_{p,i}x^\star_{p,i}
\left(
\widetilde\ell_i(p)+\beta_i-f_{C,\beta}(p)
\right)
= \sum_{p,i}x^\star_{p,i}s_{p,i},
\end{aligned}
\]

which proves \eqref{eq:kmeans-gap-equals-slack}.

The last step follows from the fact that we have,
\[
\sum_i x^\star_{p,i}=1
\qquad\text{for every }p,
\qquad\text{and}\qquad
\sum_p x^\star_{p,i}=h_i
\qquad\text{for every }i.
\]


For a ring \(R\), let
\[
\TCost_R
=
\sum_{p\in R}\sum_{i=1}^k x^\star_{p,i}\widetilde\ell_i(p),
\qquad
\Gap_R
=
\sum_{p\in R}\sum_{i=1}^k x^\star_{p,i}s_{p,i}.
\]
Thus
\[
\sum_R \TCost_R=\TCost_P(C,h),
\qquad
\sum_R \Gap_R=\Delta_{C,h}(\beta).
\]

We now bound the variance inside a ring \(R\).  Since
subtracting the anchor value does not change variance,
\[
\Var_{p\sim R}[g_R(p)]
=
\Var_{p\sim R}[f_{C,\beta}(p)]
=
\frac12
\mathbb E_{p,q\sim R}
\left(f_{C,\beta}(p)-f_{C,\beta}(q)\right)^2 .
\]
Thus it is enough to control the squared fluctuation of \(f_{C,\beta}\) on
pairs of points in \(R\).  The purpose of the next estimate is to express
this fluctuation as the sum of two terms: a truncated-cost variation term,
which will be charged to \(\TCost_R\), and a slack term, which will be
charged to \(\Gap_R\).

Fix \(p,q\in R\), and write
\[
D=\left|f_{C,\beta}(p)-f_{C,\beta}(q)\right|.
\]

First suppose that
\(f_{C,\beta}(p)\ge f_{C,\beta}(q)\). Then, using
\(s_{p,i},s_{q,i}\ge 0\), for every \(i\in[k]\),
\[
\begin{aligned}
D
&=
f_{C,\beta}(p)-f_{C,\beta}(q) \\
&=
\widetilde\ell_i(p)-\widetilde\ell_i(q)
-s_{p,i}+s_{q,i} \\
&\le
|\widetilde\ell_i(p)-\widetilde\ell_i(q)|
+
s_{q,i}.
\end{aligned}
\]
Also, by the centered range bound,
\[
D=|g_R(p)-g_R(q)|
\lesssim
\eps^{-1}\rho_R^2 .
\]
Combining these inequalities, we get, for every \(i\in[k]\),
\[
D^2
\lesssim
|\widetilde\ell_i(p)-\widetilde\ell_i(q)|^2
+
\frac{\rho_R^2}{\eps}s_{q,i}.
\]
Indeed, apply the elementary implication
\[
0\le D\le A+s,\qquad D\le B
\qquad\Longrightarrow\qquad
D^2\lesssim A^2+Bs
\]
with
\[
A=|\widetilde\ell_i(p)-\widetilde\ell_i(q)|,
\qquad
B\lesssim \eps^{-1}\rho_R^2,
\qquad
s=s_{q,i}.
\]
Averaging this inequality over \(i\) with weights \(x^\star_{q,i}\), and using
\[
\sum_i x^\star_{q,i}=1,
\]
gives
\[
D^2
\lesssim
\sum_i x^\star_{q,i}
|\widetilde\ell_i(p)-\widetilde\ell_i(q)|^2
+
\frac{\rho_R^2}{\eps}
\sum_i x^\star_{q,i}s_{q,i}.
\]

The case \(f_{C,\beta}(q)>f_{C,\beta}(p)\) is symmetric. In that case,
the same argument gives
\[
D^2
\lesssim
\sum_i x^\star_{p,i}
|\widetilde\ell_i(p)-\widetilde\ell_i(q)|^2
+
\frac{\rho_R^2}{\eps}
\sum_i x^\star_{p,i}s_{p,i}.
\]
Combining the two cases, and upper bounding  \(D^2\)
 by the sum of the two bounds, we obtain
\[
\begin{aligned}
\left(f_{C,\beta}(p)-f_{C,\beta}(q)\right)^2
\lesssim
&\sum_i
\bigl(x^\star_{p,i}+x^\star_{q,i}\bigr)
|\widetilde\ell_i(p)-\widetilde\ell_i(q)|^2 \\
&+
\frac{\rho_R^2}{\eps}
\sum_i
\bigl(x^\star_{p,i}s_{p,i}
+
x^\star_{q,i}s_{q,i}\bigr).
\end{aligned}
\]
First, we focus on the first term and bound the variation of the truncated squared costs.  We claim
that, for every \(i\) and every \(p,q\in R\),
\[
|\widetilde\ell_i(p)-\widetilde\ell_i(q)|^2
\lesssim
\rho_R^2\widetilde\ell_i(p)+\rho_R^4.
\]

If the truncated cost is constant on \(R\), then
\(\widetilde\ell_i(p)=\widetilde\ell_i(q)\), and there is nothing to prove.
Otherwise, using the triangle inequality, we get
\[
\begin{aligned}
|\widetilde\ell_i(p)-\widetilde\ell_i(q)|
&=
|d(p,c_i)^2-d(q,c_i)^2| \\
&\le d(p,q)\bigl(d(p,c_i)+d(q,c_i)\bigr) \\
&\le d(p,q)\bigl(2d(p,c_i)+d(p,q)\bigr) \\
&\lesssim \rho_R d(p,c_i)+\rho_R^2,
\end{aligned}
\]
where the last step uses \(p,q\in R\).  Squaring gives
\[
|\widetilde\ell_i(p)-\widetilde\ell_i(q)|^2
\lesssim
\rho_R^2 d(p,c_i)^2+\rho_R^4
=
\rho_R^2\widetilde\ell_i(p)+\rho_R^4,
\]
as claimed.

We now average the pairwise bound over independent uniform points
\(p,q\sim R\).  Using the truncated-cost variation bound, the cost-variation
part is at most
\[
\begin{aligned}
&\mathbb E_{p,q\sim R}
\sum_i
\bigl(x^\star_{p,i}+x^\star_{q,i}\bigr)
|\widetilde\ell_i(p)-\widetilde\ell_i(q)|^2 \\
&\qquad\lesssim
\mathbb E_{p,q\sim R}
\left[
\rho_R^2\sum_i x^\star_{p,i}\widetilde\ell_i(p)
+
\rho_R^2\sum_i x^\star_{q,i}\widetilde\ell_i(q)
+
\rho_R^4
\right] \\
&\qquad\lesssim
\frac{\rho_R^2}{n_R}\TCost_R+\rho_R^4 .
\end{aligned}
\]


Second, using the definition of \(\Gap_R\), the second term  contributes
\[
\begin{aligned}
&\mathbb E_{p,q\sim R}
\frac{\rho_R^2}{\eps}
\sum_i
\bigl(x^\star_{p,i}s_{p,i}
+
x^\star_{q,i}s_{q,i}\bigr) \\
&\qquad\lesssim
\frac{\rho_R^2}{\eps n_R}\Gap_R.
\end{aligned}
\]
Therefore,
\[
\Var_{p\sim R}[g_R(p)]
\lesssim
\frac{\rho_R^2}{n_R}\TCost_R
+
\rho_R^4
+
\frac{\rho_R^2}{\eps n_R}\Gap_R .
\]

Multiplying by \(n_R/\rho_R^2\) and summing over rings gives
\[
\begin{aligned}
\sum_R
\frac{n_R}{\rho_R^2}
\Var_{p\sim R}[g_R(p)]
&\lesssim
\sum_R \TCost_R
+
\sum_R n_R\rho_R^2
+
\eps^{-1}\sum_R \Gap_R \\
&=
\TCost_P(C,h)
+
W_2
+
\eps^{-1}\Delta_{C,h}(\beta) \\
&\lesssim
\TCost_P(C,h)
+
\eps^{-1}\Delta_{C,h}(\beta)
\end{aligned}
\]

as required.

\subsubsection{Proof of the dual-vector discretization lemma}

We prove \Cref{lem:kmeans-dual-vector-discretization}. Fix
\(C\in X^k\), and let \(\mathcal D\) be the set of points lying in
 rings. Choose a constant \(K \lesssim \eps^{-1}\) such that the
centered range bound gives
\[
\left|
f_{C,\beta}(p)-f_{C,\beta}(a_R)
\right|
\le
K\rho_R^2
\qquad
\text{for every }\beta\in\mathbb R^k
\text{ and every }p\in R .
\]
For \(p\in\mathcal D\), where \(R\) is the ring containing \(p\), define
the normalized residual directly by
\[
G_{C,\beta}(p)
=
\frac{
f_{C,\beta}(p)-f_{C,\beta}(a_R)
}{
K\rho_R^2
}.
\]
Thus \(G_{C,\beta}\in[-1,1]^{\mathcal D}\).

We use the following residual-net bound.

\begin{lemma}[Residual net for fixed centers]
\label{lem:kmeans-residual-net}
Fix \(C\in X^k\). For every \(\gamma\in(0,1)\), the class
\[
\mathcal G_C
=
\{G_{C,\beta}:\beta\in\mathbb R^k\}
\]
has a \(\gamma\)-net in \(\ell_\infty(\mathcal D)\) of size at most
\[
\left(\frac{Nk}{\gamma}\right)^{O(k)}.
\]
\end{lemma}


The lemma follows by an almost identical argument as in the proof of \Cref{lem:kmedian-residual-net}. Instead of working with \(d(p,c)\), we have truncated costs \(\widetilde{\ell}_i\). For fixed \(C\), the minimizer pattern is still
determined by an arrangement of \(O(Nk^2)\) hyperplanes in \(\beta\)-space,
and on each cell the residual vector \(G_{C,\beta}\) is affine in \(\beta\).
Since the centered-range bound gives \(G_{C,\beta}\in[-1,1]^{\mathcal D}\),
the same volumetric covering argument gives the claimed net size.

The discretization argument also uses two deterministic bounds due to
the ring decomposition. First, the allowed error \(\Allow(\beta)\) is never
smaller than \(\eps\Wtwo\). Second, the total sampling
error is always bounded by the maximum centered range on the rings.

The next lemma records two deterministic bounds that set the relevant
scale for the discretization: the allowed error is always at least the
ring scale, and the sampling error is always at most the total centered
range over the rings.

\begin{lemma}[Deterministic bound]
\label{lem:kmeans-deterministic-envelope}
For every \(\beta\in\mathbb R^k\),
\[
\Allow(\beta)\gtrsim \eps\Wtwo
\]
and
\[
|\Error_C(\beta)|
\lesssim
\eps^{-1}\Wtwo.
\]
\end{lemma}

\begin{proof}
For every feasible \loadvector \(h\), the ring construction and the
truncation comparison imply
\[
W_2 \lesssim \OPT \le \cost_P(C,h) \lesssim \widetilde{\cost}_P(C,h).
\]
Since dual gaps are nonnegative,
\[
\Allow(\beta)
=
\inf_{h\in n\Delta_k}
\left(
\eps\,\TCost_P(C,h)
+
\frac14\Delta_{C,h}(\beta)
\right)
\gtrsim
\eps\Wtwo .
\]

For the error bound, we write the sampling error in centered form.  Since
the sampled points from ring \(R\) have total weight exactly \(n_R\), the
anchor value \(f_{C,\beta}(a_R)\) cancels, and
\[
\Error_C(\beta)
=
\sum_R n_R
\left(
\frac1{m_R}\sum_{q\in Q_R}
\bigl(f_{C,\beta}(q)-f_{C,\beta}(a_R)\bigr)
-
\frac1{n_R}\sum_{p\in R}
\bigl(f_{C,\beta}(p)-f_{C,\beta}(a_R)\bigr)
\right).
\]
By the centered range bound, every term inside the averages for ring \(R\)
has magnitude at most \(O(\eps^{-1}\rho_R^2)\).  Hence the two averages
differ by at most \(O(\eps^{-1}\rho_R^2)\), and therefore
\[
|\Error_C(\beta)|
\lesssim
\sum_R n_R\eps^{-1}\rho_R^2
=
\eps^{-1}\Wtwo .
\]
\end{proof}

\begin{proof}[Proof of \Cref{lem:kmeans-dual-vector-discretization}]
By \Cref{lem:kmeans-deterministic-envelope}, every dual vector satisfies
\(\Allow(\beta)\gtrsim \eps\Wtwo\) and \({|\Error_C(\beta)|
\lesssim
\eps^{-1}\Wtwo}\). 
It remains to consider the range
\[
\eps\Wtwo
\lesssim
\Allow(\beta)
\lesssim
\eps^{-1}\Wtwo .
\]
Indeed, if \(\Allow(\beta)\) is larger than a sufficiently large constant
multiple of \(\eps^{-1}\Wtwo\), then the  bound
\(
|\Error_C(\beta)|\lesssim \eps^{-1}\Wtwo
\)
already implies \(|\Error_C(\beta)|\le \Allow(\beta)\).

\noindent
Let
\[
\lambda_0=c\eps\Wtwo,
\qquad
\lambda_j=2^j\lambda_0,
\]
and define the shells
\[
\mathcal S_j
=
\{\beta\in\mathbb R^k:\lambda_j\le \Allow(\beta)<2\lambda_j\}.
\]
Only \(O(\log(1/\eps))\) shells can intersect the relevant range above.

For each nonempty shell \(\mathcal S_j\), take a net for
\(\{G_{C,\beta}:\beta\in\mathcal S_j\}\) in the metric
\(\ell_\infty(\mathcal D)\) at resolution
\[
\gamma=c_0\eps^2,
\]
where \(c_0>0\) is a sufficiently small universal constant.  By
\Cref{lem:kmeans-residual-net}, each shell has a net of size at most
\[
\left(\frac{Nk}{\eps}\right)^{O(k)}.
\]
Taking the union over all relevant shells gives a representative set
\(\mathcal B_C\) satisfying
\[
|\mathcal B_C|
\le
\exp\{O(k\log(N/\eps))\}.
\]

Conditioning on the event that every representative
\(\beta'\in\mathcal B_C\) satisfies
\(
|\Error_C(\beta')|\le \frac18 \Allow(\beta'),
\)
we show that the same type of bound holds for every
\(\beta\in\mathbb R^k\) in the relevant range.  Fix such a \(\beta\), and
let \(j\) be such that \(\beta\in\mathcal S_j\).  Choose
\(\beta'\in\mathcal B_C\cap\mathcal S_j\) such that
\[
\|G_{C,\beta}-G_{C,\beta'}\|_{\infty}
\le
\gamma .
\]
Since \(\beta'\in\mathcal S_j\), we have
\[
\Allow(\beta')<2\lambda_j,
\]
and therefore
\[
|\Error_C(\beta')|
\le
\frac18 \Allow(\beta')
<
\frac14\lambda_j .
\]

It remains to compare the errors of \(\beta\) and \(\beta'\).  Using the
centered form of the sampling error, the anchor terms cancel ring by ring.
By the definition of \(G_{C,\beta}\),
\[
\begin{aligned}
|\Error_C(\beta)-\Error_C(\beta')|
&\lesssim
\eps^{-1}\Wtwo
\|G_{C,\beta}-G_{C,\beta'}\|_{\infty} \\
&\le
\eps^{-1}\Wtwo \gamma \\
&=
c_0\eps\Wtwo .
\end{aligned}
\]
By the lower bound on the allowed error,
\(\Allow(\beta)\gtrsim \eps\Wtwo\).  Since \(\beta\in\mathcal S_j\), this also
means \(\lambda_j\gtrsim \eps\Wtwo\).  Choosing \(c_0\) sufficiently small,
we get
\[
|\Error_C(\beta)-\Error_C(\beta')|
\le
\frac14\lambda_j .
\]

\noindent
Combining the two estimates,
\[
|\Error_C(\beta)|
\le
|\Error_C(\beta')|
+
|\Error_C(\beta)-\Error_C(\beta')|
<
\frac12\lambda_j
\le
\Allow(\beta).
\]
This proves \Cref{lem:kmeans-dual-vector-discretization}.
\end{proof}

\section{Euclidean Capacitated \texorpdfstring{\(\kmedian\)}{k-median}}
\label{sec:euclidean-kmedian}

We next prove the Euclidean coreset theorem for capacitated \(\kmedian\). Note that, in this section, we reuse the variable \(d\) to denote the dimension of the Euclidean space the points lie in.

\begin{theorem}[Euclidean capacitated \(\kmedian\)]
\label{thm:euclidean-kmedian}
Let \(P\subseteq \mathbb R^d\) be a set of $n$ points. For every \(\eps,\delta\in(0,1)\), there is an
algorithm which outputs a weighted subset \(Q\subseteq P\) of size
\[
|Q|
=
\widetilde O\left(k\eps^{-4}+k\eps^{-2}\log(1/\delta)
\right)
\]
such that, with probability at least \(1-\delta\), \(Q\) is an
\(\eps\)-coreset for Euclidean capacitated \(\kmedian\).
\end{theorem}

We will in fact show the following dimension-dependent coreset bound.
\begin{lemma}[Dimension-dependent coreset]
\label{lem:euclidean-kmedian-dimension-dependent}
There is an algorithm which, with probability at least \(1-\delta\),
outputs an \(\eps\)-coreset \(Q\subseteq P\) for Euclidean capacitated
\(\kmedian\) of size
\[
\widetilde O\!\left(kd\eps^{-2}+k\eps^{-2}\log(1/\delta)\right).
\]
\end{lemma}
Once \Cref{lem:euclidean-kmedian-dimension-dependent} is proved,
\Cref{thm:euclidean-kmedian} follows via the ideas of terminal embeddings and the
existence of \(\poly(k/\eps)\)-size coresets for Euclidean $\kmedian$.  This reduction is
standard, so we defer the details to
\Cref{sec:dimension-independent-bound}.

\subsection{Mass-Preserving Sensitivity Sampling}
We set up some basic notation required to state the algorithm.

Let \(A=\{a_1,\ldots,a_k\}\subseteq \mathbb R^d\) be a constant-factor
approximate unconstrained \(\kmedian\) solution for \(P\). For $j \in [k]$, let $P_j$ be the set of points in $P$ whose nearest anchor in $A$ is $a_j$ with ties being broken arbitrarily.  For each \(p\in P_j\), define
\(
r_p=\|p-a_j\|
\) and let
\[
\Phi_j=\sum_{p\in P_j} r_p,
\qquad
\Wone
=
\sum_{j \in [k]} \Phi_j
\lesssim \opt ,
\]
where \(\opt\) denotes the optimum unconstrained Euclidean \(\kmedian\)
cost of \(P\).

\paragraph{Algorithm.}
For each cluster $P_j$, the algorithm samples $m_j$
points from it where $m_j$ is chosen proportional to the cost $\Phi_j$ of the cluster. Within \(P_j\), the samples are drawn i.i.d. from a distribution \(\pi_j\)
which is a mixture of the uniform distribution and the distribution
proportional to \(r_p\). Sampling uniformly ensures that the mass of the distribution is preserved up to a $(1\pm \eps)$ factor and sampling high-cost points with larger probability ensures that the variance of the estimator is low.  After sampling, the weights of
points within the cluster are rescaled so that the total weight of the
samples from cluster \(P_j\) is exactly \(|P_j|\).

Throughout we assume that each cluster $P_j$ has cost $\Phi_j > 0$. This is without loss of generality: if a cluster has a cost $0$, then it can be represented by a single point with weight equal to the size of the cluster. This increases the coreset size by at most $k$.

\begin{algorithm}[H]
\caption{Mass-preserving sensitivity sampling for Euclidean capacitated \(\kmedian\)}
\label{alg:euclidean-kmedian-mass-preserving}
\begin{algorithmic}[1]
\State Compute the $O(1)$-approximate solution $A$ and the clusters  $P_1, \ldots, P_k$ induced by $A$. 
\State Let \(T=\widetilde\Theta(k\eps^{-2}(d+\log(1/\delta)))\) and
\(m_0=\Theta(\eps^{-2}\log(k/\delta))\).
\For{each cluster \(P_j\)}
    \State Set \(m_j=\left\lceil T\Phi_j/\Wone\right\rceil+m_0\). Sample
    a multiset \(Q_j\) of \(m_j\) points i.i.d. from the distribution
    \[ \pi_j(p)=\frac12\left(\frac1{|P_j|}+\frac{r_p}{\Phi_j}\right),
    \qquad p\in P_j .
    \]
    \State Assign each occurrence \(q\in Q_j\) the initial weight 
    \(\widehat w_q=(m_j\pi_j(q))^{-1}\).
    \State Let
    \(  \lambda_j=|P_j|/\left({\sum_{q\in Q_j}\widehat w_q} \right).
    \)
    \State Rescale the weights, setting
    \(w_q=\lambda_j\widehat w_q\) for every \(q\in Q_j\), so that $\sum_{q \in Q_j} w_q = |P_j|$.
\EndFor
\State \textbf{Output:} the resulting weighted multiset
\(Q=\bigcup_{j=1}^k Q_j\).
\end{algorithmic}
\end{algorithm}
We note some simple properties of the algorithm that will be used in the analysis.
\begin{claim}[Cluster Mass Preservation]
\label{clm:cluster-mass}
The following properties hold.
\begin{enumerate}[label=(\roman*)]
    \item \emph{\textbf{(Initial Weights) }}With probability at least \(1-\delta\), for all $j \in [k]$, we have 
    \(\sum_{q\in Q_j}\widehat w_q=(1\pm\eps)|P_j|\).
    \item \emph{\textbf{(Final Weights) }} With probability \(1\), for all $j \in [k]$, we have 
    \(\sum_{q\in Q_j}w_q=|P_j|\).
    \item If the event in (i) occurs, then for all  $q \in Q$ we have $|w_q - \widehat w_q| \lesssim \eps  \widehat{w}_q$. 
\end{enumerate}
\end{claim}

\begin{proof}
(i) Fix \(j\in[k]\), and write
\(\widehat M_j=\sum_{q\in Q_j}\widehat w_q\). Each sampled occurrence has
expected initial weight \(|P_j|/m_j\), so
\(\mathbb E[\widehat M_j]=|P_j|\). Moreover, since
\(\pi_j(p)\ge 1/(2|P_j|)\), every initial weight is at most
\(2|P_j|/m_j\). Hence Hoeffding's inequality (see \ref{lem:hoeffding}) gives
\[
\Pr\bigl[|\widehat M_j-|P_j||>\eps |P_j|\bigr]
\le
2\exp(-\Omega(\eps^2m_j))
\le
\delta/k,
\]
where we used
\(m_j\ge m_0=\Theta(\eps^{-2}\log(k/\delta))\). A union bound over
\(j\in[k]\) proves (i).

(ii) This  follows immediately from the rescaling step of the algorithm.  

(iii) It follows from (i), (ii) that the rescaling factor \(\lambda_j=1\pm O(\eps)\), which implies (iii). 
\end{proof}

\subsection{Analysis}

Our main goal is to prove
\Cref{lem:euclidean-kmedian-dimension-dependent}. We follow the same
dual-based approach used in \Cref{sec:finite-kmedian} for finite-metric
\(\kmedian\). The quantities
\(f_{C,\beta}\), \(\DualCost_S(C,h,\beta)\), and \(\Error_C(\beta)\) are
used with the same meaning as before, with Euclidean distance replacing
the finite metric.

The following lemma is the same dual-to-primal step proved in the finite-metric
section. 
\begin{restatable}[Dual preservation implies load preservation]{lemma}{EuclideanDualToPrimal}
\label{lem:euclidean-kmedian-dual-to-primal}
For every fixed \(C\in(\mathbb R^d)^k\), if
\begin{align}
\sup_{\beta\in\mathbb R^k}|\Error_C(\beta)|\le \eta, \quad \text{then} \quad 
\sup_{h\in n\Delta_k}
|\cost_Q(C,h)-\cost_P(C,h)|
\le \eta .
\end{align}
\end{restatable}
\noindent We next center the dual functions $f_{C,\beta}$ at the anchors. For each point $p$ that is in cluster $P_j$ 
\begin{align}
g_{C,\beta}(p)=f_{C,\beta}(p)-f_{C,\beta}(a_j).
\end{align}
Since \(f_{C,\beta}\) is 1-Lipschitz for \(\kmedian\),
\begin{equation}
\label{eq:g-bound}
|g_{C,\beta}(p)|\le \|p-a_j\|=r_p .
\end{equation}
Furthermore, we can express the dual error in terms of $g_{C,\beta}$. Here we crucially use that the algorithm preserves the mass of each cluster. 

\begin{claim}
\label{clm:euclidean-kmedian-anchor-cancellation}
For every \(C\) and \(\beta\),
\(\Error_C(\beta)
=
\sum_{q\in Q}w_qg_{C,\beta}(q)
-
\sum_{p\in P}g_{C,\beta}(p).\)
\end{claim}

\begin{proof}
Fix \(j \in  [k]\). Since
\(f_{C,\beta}(p)=g_{C,\beta}(p)+f_{C,\beta}(a_j)\) for a point $p \in P_j$,
\[
\begin{aligned}
\sum_{q\in Q_j}w_qf_{C,\beta}(q)
-
\sum_{p\in P_j}f_{C,\beta}(p)
&=
\sum_{q\in Q_j}w_qg_{C,\beta}(q)
-
\sum_{p\in P_j}g_{C,\beta}(p) 
+
f_{C,\beta}(a_j)
\left(
\sum_{q\in Q_j}w_q-|P_j|
\right)\\
&=\sum_{q\in Q_j}w_qg_{C,\beta}(q)
-
\sum_{p\in P_j}g_{C,\beta}(p).
\end{aligned}
\]
The last term is zero by property (ii) of \Cref{clm:cluster-mass}.
Summing over \(j\in[k]\) proves the claim.
\end{proof}

Due to the rescaling of weights that the algorithm does, the right-hand side of the identity in \Cref{clm:euclidean-kmedian-anchor-cancellation}, which involves the final weights $w_q$, is not a sum of independent random variables. In order to be able to apply concentration inequalities, we decompose
the error into terms as follows and bound each of them separately. 
\begin{equation}
\label{eq:euclidean-kmedian-rescaling-decomposition}
|\Error_C(\beta)|
\le
\left|
\sum_{q\in Q}\widehat w_qg_{C,\beta}(q)
-
\sum_{p\in P}g_{C,\beta}(p)
\right|
+
\left|
\sum_{q\in Q}(w_q-\widehat w_q)g_{C,\beta}(q)
\right|.
\end{equation}
\begin{restatable}[Error before rescaling]{lemma}{EuclideanKMedianBeforeRescaling}
\label{lem:euclidean-kmedian-before-rescaling}
With probability at least \(1-\delta/2\),
\[
\sup_{C,\beta}
\left|
\sum_{q\in Q}\widehat w_qg_{C,\beta}(q)
-
\sum_{p\in P}g_{C,\beta}(p)
\right|
\lesssim
\eps \Wone .
\]
\end{restatable}

\begin{restatable}[Rescaling error]{lemma}{EuclideanKMedianRescalingError}
\label{lem:euclidean-kmedian-rescaling-error}
With probability at least \(1-\delta/2\),
\[
\sup_{C,\beta}
\left|
\sum_{q\in Q}(w_q-\widehat w_q)g_{C,\beta}(q)
\right|
\lesssim
\eps \Wone .
\]
\end{restatable}

We now show that the above lemmas complete the proof of the coreset bound.

\begin{proof}[Proof of \Cref{lem:euclidean-kmedian-dimension-dependent}]
By \Cref{lem:euclidean-kmedian-before-rescaling},
\Cref{lem:euclidean-kmedian-rescaling-error}, and
\eqref{eq:euclidean-kmedian-rescaling-decomposition}, with probability at
least \(1-\delta\),
\[
\sup_{C,\beta}|\Error_C(\beta)|
\lesssim
\eps \Wone .
\]
Thus, by \Cref{lem:euclidean-kmedian-dual-to-primal}, for every
\(C\in(\mathbb R^d)^k\) and \(h\in n\Delta_k\),
\[
|\cost_Q(C,h)-\cost_P(C,h)|
\lesssim
\eps \Wone
\lesssim
\eps \opt
\lesssim
\eps \cost_P(C,h).
\]
The bound on the coreset size  follows from the choice of \(m_j\):
\[
\sum_{j \in  [k]}m_j
\le
T+k+km_0
=
\widetilde O\!\left(kd\eps^{-2}+k\eps^{-2}\log(1/\delta)\right).
\]
\end{proof}
The rest
of this section is devoted to proving
\Cref{lem:euclidean-kmedian-before-rescaling} and
\Cref{lem:euclidean-kmedian-rescaling-error}.

\subsection{Bounding the Error Before Rescaling Via Pseudo-Dimension:
Proof of \Cref{lem:euclidean-kmedian-before-rescaling}}

We prove \Cref{lem:euclidean-kmedian-before-rescaling} using the
pseudo-dimension tools collected in \Cref{app:pseudodimension-tools}. The
argument is to normalize the sampled summands, bound the
pseudo-dimension of the resulting function class, and then apply uniform
convergence.

For each set of centers \(C\) and dual vector \(\beta\), define
\begin{equation}
\label{eq:euclidean-kmedian-F-definition}
F_{C,\beta}(p)
=
\frac{T}{2\Wone}\widehat w_p g_{C,\beta}(p).
\end{equation}
In the above equation, for a point $p$ in cluster $P_j$ we use  $\widehat w_p =1/(m_j\pi_j(p))$. This is the initial weight assigned if  $p$ is sampled by   
\Cref{alg:euclidean-kmedian-mass-preserving}. The point of defining $F$ is that we can express the error before weight rescaling in terms of $F$:
\begin{equation}
\label{eq:euclidean-kmedian-before-rescaling-F}
\sum_{q\in Q}\widehat w_qg_{C,\beta}(q)
-
\sum_{p\in P}g_{C,\beta}(p)
=
\frac{2\Wone}{T}
\left(
\sum_{q\in Q} F_{C,\beta}(q)
-
\sum_{q\in Q} \mathbb E F_{C,\beta}(q)
\right).
\end{equation}
The normalization factor \(T/(2\Wone)\) in \eqref{eq:euclidean-kmedian-F-definition} is chosen so that \(F_{C,\beta}\) is bounded by
\(1\), as shown below.

\begin{claim}
\label{clm:euclidean-kmedian-F-bounded}
For every \(C,\beta\), and every \(p\in P\),
\(
|F_{C,\beta}(p)|\le 1
\).
\end{claim}

\begin{proof}
Let \(p\in P_j\). If \(r_p=0\), then \(g_{C,\beta}(p)=0\), so the claim
is immediate. Otherwise, \(|g_{C,\beta}(p)|\le r_p\),
\(\pi_j(p)\ge r_p/(2\Phi_j)\), and \(m_j\ge T\Phi_j/\Wone\). Hence
\[
|F_{C,\beta}(p)|
=
\frac{T}{2\Wone}
\frac{|g_{C,\beta}(p)|}{m_j\pi_j(p)}
\le 1 . 
\]
\end{proof}
We now bound the pseudo-dimension of the following function class:
\begin{equation}
\label{eq:euclidean-kmedian-F-class}
\mathcal F
=
\{F_{C,\beta}:C\in(\mathbb R^d)^k,\ \beta\in\mathbb R^k\}.
\end{equation}

\begin{lemma}
\label{lem:euclidean-kmedian-summand-pdim}
\(\pdim(\mathcal F)=\widetilde O(kd).\)
\end{lemma}

\begin{proof}
Fix points \(p_1,\ldots,p_m\in P\) and thresholds
\(\tau_1,\ldots,\tau_m\in\mathbb R\). We bound the number of possible
sign patterns of \(F_{C,\beta}(p_\ell)-\tau_\ell\) as \(C\) and \(\beta\)
vary.

For \(p_\ell\in P_{j_\ell}\), set
\[
\alpha_\ell=\frac{T}{2\Wone}\widehat w_{p_\ell}>0 .
\]
Then \(F_{C,\beta}(p_\ell)\ge \tau_\ell\) is equivalent to
\begin{align}
f_{C,\beta}(p_\ell)-f_{C,\beta}(a_{j_\ell})
\ge
\tau_\ell/\alpha_\ell.  
\end{align}
Recall that
\(f_{C,\beta}(x)=\min_{t\in[k]}\{\|x-c_t\|+\beta_t\}\). 
Thus each threshold sign is determined by the minimizers defining
\(f_{C,\beta}(p_\ell)\) and \(f_{C,\beta}(a_{j_\ell})\), together with one
comparison between the corresponding chosen values.

All relevant comparisons that arise have the form
\begin{align}\label{eq:comparison}
\|x-c_s\|+\beta_s
\le
\|y-c_t\|+\beta_t+\theta,
\end{align}
where \(x,y,\theta\) are fixed. Writing
\(A=\beta_t+\theta-\beta_s\), \(u=\|x-c_s\|^2\), and
\(v=\|y-c_t\|^2\), the comparison in \eqref{eq:comparison} is \(\sqrt u\le \sqrt v+A\). Its truth
value is determined by the signs of
\[
A,\qquad
v-A^2,\qquad
u-v-A^2,\qquad
(u-v-A^2)^2-4A^2v .
\]
Indeed, the only point to be careful about is that one cannot square an
inequality unless the relevant side is known to be nonnegative. The signs
of \(A\) and \(v-A^2\) handle these case distinctions. Once the valid
squaring step is determined, the signs of \(u-v-A^2\) and
\((u-v-A^2)^2-4A^2v\) decide the remaining comparison.

These polynomials have degree at most \(4\) in \((C,\beta)\). There are
\(s=O(kd)\) real variables. For each point \(p_\ell\), the sign of
\(F_{C,\beta}(p_\ell)-\tau_\ell\) is determined by the signs of
\(O(k^2)\) constant-degree polynomials: these account for the possible
minimizers at \(p_\ell\) and at its anchor, and for the comparison between
the two chosen values.

Thus, over all \(m\) points, the threshold signs are determined by
\(N_{\rm poly}=O(mk^2)\) constant-degree polynomials. Warren's theorem
implies that the number of possible threshold patterns is at most
\[
\left(\frac{C_0N_{\rm poly}}{s}\right)^s .
\]
If the \(m\) points are pseudo-shattered, then this number is at least
\(2^m\). Since \(s=O(kd)\), this gives \(m=\widetilde O(kd)\).
\end{proof}

\EuclideanKMedianBeforeRescaling*
\begin{proof}
Let \(M\) be the total number of points sampled by the algorithm and let 
\(D=\pdim(\mathcal F)\). By \Cref{lem:uniform-convergence}, applied to
\(\mathcal F\), together with
\Cref{lem:euclidean-kmedian-summand-pdim} and
\Cref{clm:euclidean-kmedian-F-bounded}, with probability at least
\(1-\delta/2\),
\[
\sup_{C,\beta}
\left|
\frac1M\sum_{q\in Q}F_{C,\beta}(q)
-
\frac1M\sum_{q\in Q}\mathbb E F_{C,\beta}(q)
\right|
\lesssim
\!
\sqrt{\frac{D\log M+\log(1/\delta)}{M}}.
\]
Using \Cref{eq:euclidean-kmedian-before-rescaling-F}, this gives
\[
\sup_{C,\beta}
\left|
\sum_{q\in Q}\widehat w_qg_{C,\beta}(q)
-
\sum_{p\in P}g_{C,\beta}(p)
\right|
\lesssim
\frac{\Wone}{T}
\sqrt{M\bigl(D\log M+\log(1/\delta)\bigr)}.
\]
Since \(D=\widetilde O(kd)\), \(M=\widetilde O(T)\), and
\(T=\widetilde\Theta(k\eps^{-2}(d+\log(1/\delta)))\), the RHS
is at most \(O(\eps)\Wone\).
\end{proof}

\subsection{Bounding Rescaling Error: Proof of \Cref{lem:euclidean-kmedian-rescaling-error}}

\EuclideanKMedianRescalingError*

\begin{proof}
Fix \(j \in [k]\). By the triangle inequality,
\begin{align}
\label{eq:initial-final-diff}
\left|
\sum_{q\in Q_j}(w_q-\widehat w_q)g_{C,\beta}(q)
\right|
\le
\sum_{q\in Q_j}|w_q-\widehat w_q|\,|g_{C,\beta}(q)| 
\lesssim
\eps\sum_{q\in Q_j}\widehat w_q r_q .
\end{align}
The last inequality uses property (iii) of
\Cref{clm:cluster-mass} and the bound \(|g_{C,\beta}(q)|\le r_q\) from
\Cref{eq:g-bound}.
For every sampled point \(q\in P_j\),
\begin{align*}
\widehat w_q r_q
=
\frac{r_q}{m_j\pi_j(q)}
\le
\frac{2\Phi_j}{m_j},
\end{align*}
because \(\pi_j(q)\ge r_q/(2\Phi_j)\). Therefore
\begin{align}\label{eq:sum-weight-distance}
\sum_{q\in Q_j}\widehat w_q r_q
\le
2\Phi_j .
\end{align}
Summing \eqref{eq:initial-final-diff} over all $j \in [k]$ and using \eqref{eq:sum-weight-distance} yields,
\[
\left|
\sum_{q\in Q}(w_q-\widehat w_q)g_{C,\beta}(q)
\right|
\lesssim
\eps \sum_{j \in  [k]}\Phi_j \lesssim \eps \Wone, 
\]
which completes the proof. 
\end{proof}

\section{Euclidean Capacitated \texorpdfstring{\(\kmeans\)}{k-means}}
\label{sec:euclidean-kmeans}

We next prove the Euclidean coreset theorem for capacitated \(\kmeans\).

\begin{theorem}[Euclidean capacitated \(\kmeans\)]
\label{thm:euclidean-kmeans}
Let \(P\subseteq \mathbb R^d\) be a set of \(n\) points. For every
\(\eps,\delta\in(0,1)\), there is an algorithm which outputs a weighted
subset \(Q\subseteq P\) of size
\[
|Q|
=
\widetilde O\left(k\eps^{-6}+k\eps^{-4}\log(1/\delta)\right)
\]
such that, with probability at least \(1-\delta\), \(Q\) is an
\(\eps\)-coreset for Euclidean capacitated \(\kmeans\).
\end{theorem}

As in the Euclidean \(\kmedian\) section, it is enough to prove a
dimension-dependent coreset bound. The reduction from the
dimension-dependent bound to \Cref{thm:euclidean-kmeans} is in \Cref{sec:dimension-independent-bound}.

\begin{lemma}[Dimension-dependent coreset]
\label{lem:euclidean-kmeans-dimension-dependent}
There is an algorithm, which with probability
at least \(1-\delta\) outputs an \(\eps\)-coreset \(Q\subseteq P\) for
Euclidean capacitated \(\kmeans\) of size
\[
\widetilde O\!\left(kd\eps^{-4}+k\eps^{-4}\log(1/\delta)\right).
\]
\end{lemma}

\subsection{Mass-Preserving Sensitivity Sampling}

We use the same reference clusters as in the Euclidean \(\kmedian\)
section, but with a squared reference scale. Let
\(A=\{a_1,\ldots,a_k\}\subseteq\mathbb R^d\) be an \(O(1)\)-approximate
unconstrained Euclidean \(\kmeans\) solution for \(P\). For \(j\in[k]\),
let \(P_j\) be the set of points in \(P\) whose nearest anchor in \(A\) is
\(a_j\), with ties broken arbitrarily. For each \(p\in P_j\), define
\(r_p=\|p-a_j\|\), and let
\[
\Phi_j=\sum_{p\in P_j}r_p^2,
\qquad
\Wtwo=\sum_{j\in[k]}\Phi_j\lesssim \opt ,
\]
where \(\opt\) denotes the optimum unconstrained Euclidean \(\kmeans\)
cost of \(P\).

The sampler is the same as in
\Cref{alg:euclidean-kmedian-mass-preserving}, with \(r_p^2\) replacing
\(r_p\). Thus, for every cluster \(P_j\) with \(\Phi_j>0\), we sample a
multiset \(Q_j\) of
\(
m_j=\left\lceil T\Phi_j/\Wtwo\right\rceil+m_0
\)
points independently from
\[
\pi_j(p)
=
\frac12\left(\frac1{|P_j|}+\frac{r_p^2}{\Phi_j}\right),
\qquad p\in P_j,
\]
where \(T=\widetilde\Theta(k\eps^{-4}(d+\log(1/\delta)))\) and
\(m_0=\Theta(\eps^{-4}\log(k/\delta))\).

Each sampled occurrence \(q\in Q_j\) receives initial weight
\(\widehat w_q=(m_j\pi_j(q))^{-1}\). The weights inside each cluster are
then rescaled so that the final total weight of the samples from \(P_j\)
is exactly \(|P_j|\). We use $w_q$ to denote the weight of the point after rescaling.

We note the analogue of \Cref{clm:cluster-mass}, with essentially the
same proof. The initial weights preserve each cluster size up to a
\((1\pm\eps^2)\) factor, which will be important when bounding the
rescaling error. This stronger concentration follows from the choice
\(m_0=\Theta(\eps^{-4}\log(k/\delta))\), which ensures that sufficiently
many samples are drawn from each cluster.

\begin{claim}[Cluster Mass Preservation for \(\kmeans\)]
\label{clm:euclidean-kmeans-cluster-mass}
The following properties hold.
\begin{enumerate}[label=(\roman*)]
    \item \emph{\textbf{(Initial Weights)}} With probability at least
    \(1-\delta\), for all \(j\in[k]\),
    \(\sum_{q\in Q_j}\widehat w_q=(1\pm\eps^2)|P_j|\).
    \item \emph{\textbf{(Final Weights)}} With probability \(1\), for all
    \(j\in[k]\), \(\sum_{q\in Q_j}w_q=|P_j|\).
    \item If the event in (i) occurs, then for all \(q\in Q\),
    \(|w_q-\widehat w_q|\lesssim \eps^2\widehat w_q\).
\end{enumerate}
\end{claim}
\subsection{Analysis}

Our goal is to prove \Cref{lem:euclidean-kmeans-dimension-dependent}. The
argument is the same as for Euclidean \(\kmedian\), except that we first
truncate far point-center costs.

\paragraph{Truncation.}
For \(p\in P_j\), define the truncated squared cost of assigning \(p\) to
center \(c_i\) by
\[
\widetilde \ell_i(p)
=
\begin{cases}
\|p-c_i\|^2, & \text{if } \|a_j-c_i\|\le r_p/\eps,\\
\|a_j-c_i\|^2, & \text{otherwise}.
\end{cases}
\]
For a weighted instance \(S\) supported on \(P\), let
\(\TCost_S(C,h)\) denote the cost using the truncated costs
\(\widetilde\ell_i\). By the same pointwise argument as in
\Cref{lem:kmeans-truncation-comparison}, for every such \(S\), every
\(C\in(\mathbb R^d)^k\), and every \(h\in n\Delta_k\),
\begin{equation}
\label{eq:euclidean-kmeans-truncation-comparison}
\TCost_S(C,h)
=
(1\pm O(\eps))\cost_S(C,h).
\end{equation}

\paragraph{Centered truncated dual error.}
For the rest of the analysis, the dual quantities
\(f_{C,\beta}\), \(\DualCost_S(C,h,\beta)\), and \(\Error_C(\beta)\) are
interpreted with respect to the truncated edge costs
\(\widetilde\ell_i\). When evaluating \(f_{C,\beta}\) at an anchor
\(a_j\), the edge cost to \(c_i\) is \(\|a_j-c_i\|^2\).

For \(p\in P_j\), define
\[
g_{C,\beta}(p)
=
f_{C,\beta}(p)-f_{C,\beta}(a_j).
\]
The centered-range bound from the finite-metric \(\kmeans\) analysis gives
\begin{equation}
\label{eq:euclidean-kmeans-g-range}
|g_{C,\beta}(p)|\lesssim \eps^{-1}r_p^2 .
\end{equation}

By the same dual-to-primal implication as
\Cref{lem:euclidean-kmedian-dual-to-primal}, applied to the truncated
costs, it is enough to prove
\begin{equation}
\label{eq:euclidean-kmeans-uniform-truncated-dual}
\sup_{C,\beta}|\Error_C(\beta)|
\lesssim
\eps\Wtwo .
\end{equation}
Indeed, this gives
\[
|\TCost_Q(C,h)-\TCost_P(C,h)|
\lesssim
\eps\Wtwo
\lesssim
\eps\cost_P(C,h),
\]
using \(\Wtwo\lesssim\opt\le \cost_P(C,h)\), and then
\eqref{eq:euclidean-kmeans-truncation-comparison} transfers the guarantee
from \(\TCost\) to \(\cost\). As in the Euclidean \(\kmedian\) analysis
(see \Cref{clm:euclidean-kmedian-anchor-cancellation}),
cluster-wise mass preservation (\Cref{clm:euclidean-kmeans-cluster-mass}) implies
\begin{equation}
\label{eq:euclidean-kmeans-anchor-cancellation}
\Error_C(\beta)
=
\sum_{q\in Q}w_qg_{C,\beta}(q)
-
\sum_{p\in P}g_{C,\beta}(p).
\end{equation}
Combining this identity with the same initial/final weight decomposition
as in \Cref{eq:euclidean-kmedian-rescaling-decomposition}, it is enough to
prove the following two estimates.

\begin{restatable}[Error before rescaling]{lemma}{EuclideanKMeansBeforeRescaling}
\label{lem:euclidean-kmeans-before-rescaling}
With probability at least \(1-\delta/2\),
\[
\sup_{C,\beta}
\left|
\sum_{q\in Q}\widehat w_qg_{C,\beta}(q)
-
\sum_{p\in P}g_{C,\beta}(p)
\right|
\lesssim
\eps\Wtwo .
\]
\end{restatable}

\begin{restatable}[Rescaling error]{lemma}{EuclideanKMeansRescalingError}
\label{lem:euclidean-kmeans-rescaling-error}
With probability at least \(1-\delta/2\),
\[
\sup_{C,\beta}
\left|
\sum_{q\in Q}(w_q-\widehat w_q)g_{C,\beta}(q)
\right|
\lesssim
\eps\Wtwo .
\]
\end{restatable}

Together, the two estimates imply
\eqref{eq:euclidean-kmeans-uniform-truncated-dual} with probability at
least \(1-\delta\). By the reduction above, this gives the coreset
guarantee of \Cref{lem:euclidean-kmeans-dimension-dependent} for Euclidean
\(\kmeans\). Thus it remains only to prove
\Cref{lem:euclidean-kmeans-before-rescaling} and
\Cref{lem:euclidean-kmeans-rescaling-error}.

\subsection{Bounding the Error Before Rescaling: Proof of
\Cref{lem:euclidean-kmeans-before-rescaling}}

For each set of centers \(C\) and dual vector \(\beta\), define the function
\begin{equation}
\label{eq:euclidean-kmeans-F-definition}
F_{C,\beta}(p)
=
\frac{\eps T}{C_0\Wtwo}\widehat w_p g_{C,\beta}(p),
\end{equation}
where \(C_0\) is a sufficiently large universal constant. In the above
equation, for a point \(p\) in cluster \(P_j\) we use
\(\widehat w_p=1/(m_j\pi_j(p))\). This is the initial weight assigned if
\(p\) is sampled. The point of defining \(F\) is that we can express the
error before weight rescaling in terms of \(F\):
\begin{equation}
\label{eq:euclidean-kmeans-before-rescaling-F}
\sum_{q\in Q}\widehat w_qg_{C,\beta}(q)
-
\sum_{p\in P}g_{C,\beta}(p)
=
\frac{C_0\Wtwo}{\eps T}
\left(
\sum_{q\in Q}F_{C,\beta}(q)
-
\sum_{q\in Q}\mathbb E F_{C,\beta}(q)
\right).
\end{equation}
The normalization factor \(\eps T/(C_0\Wtwo)\) in
\eqref{eq:euclidean-kmeans-F-definition} is chosen so that
\(F_{C,\beta}\) is bounded by \(1\).

\begin{claim}
\label{clm:euclidean-kmeans-F-bounded}
For every \(C,\beta\), and every \(p\in P\),
\(
|F_{C,\beta}(p)|\le 1
\).
\end{claim}

Indeed, this follows from the proof of
\Cref{clm:euclidean-kmedian-F-bounded}, using
\Cref{eq:euclidean-kmeans-g-range},
\(\pi_j(p)\ge r_p^2/(2\Phi_j)\), and \(m_j\ge T\Phi_j/\Wtwo\), with
\(C_0\) chosen sufficiently large.

We now bound the pseudo-dimension of the following function class:
\begin{equation}
\label{eq:euclidean-kmeans-F-class}
\mathcal F
=
\{F_{C,\beta}:C\in(\mathbb R^d)^k,\ \beta\in\mathbb R^k\}.
\end{equation}

\begin{lemma}
\label{lem:euclidean-kmeans-summand-pdim}
\(
\pdim(\mathcal F)=\widetilde O(kd).
\)
\end{lemma}

\begin{proof}
This is the same Warren-theorem argument as in
\Cref{lem:euclidean-kmedian-summand-pdim}. The pointwise factor in
\eqref{eq:euclidean-kmeans-F-definition} is fixed once \(p\) is fixed, so
it does not affect the sign-pattern argument.

The only difference is that the threshold comparisons now involve
truncated squared costs. These comparisons are determined by squared-cost
inequalities
\[
\|x-c_s\|^2+\beta_s
\le
\|y-c_t\|^2+\beta_t+\theta
\]
and near/far predicates
\[
\|a_j-c_s\|^2\le r_p^2/\eps^2.
\]
All of these are constant-degree polynomial inequalities in \((C,\beta)\).
Thus, for \(m\) fixed points and thresholds, all threshold signs are
determined by \(O(mk^2)\) constant-degree polynomials in \(O(k(d+1))\)
variables. Warren's theorem gives \(\pdim(\mathcal F)=\widetilde O(kd)\).
\end{proof}

\EuclideanKMeansBeforeRescaling*

\begin{proof}
Let \(M\) be the total number of points sampled by the algorithm and let
\(D=\pdim(\mathcal F)\). By \Cref{lem:uniform-convergence}, applied to
\(\mathcal F\), together with
\Cref{clm:euclidean-kmeans-F-bounded},
\Cref{lem:euclidean-kmeans-summand-pdim}, and
\Cref{eq:euclidean-kmeans-before-rescaling-F}, with probability at least
\(1-\delta/2\),
\[
\sup_{C,\beta}
\left|
\sum_{q\in Q}\widehat w_qg_{C,\beta}(q)
-
\sum_{p\in P}g_{C,\beta}(p)
\right|
\lesssim
\frac{\Wtwo}{\eps T}
\sqrt{M\bigl(D\log M+\log(1/\delta)\bigr)} \lesssim  \eps\Wtwo.
\]
as \(D=\widetilde O(kd)\), \(M=\widetilde O(T)\), and
\(T=\widetilde\Theta(k\eps^{-4}(d+\log(1/\delta)))\). 
\end{proof}
\subsection{Bounding Rescaling Error: Proof of
\Cref{lem:euclidean-kmeans-rescaling-error}}

\EuclideanKMeansRescalingError*

\begin{proof}
This is the same argument as in the Euclidean \(\kmedian\) rescaling
proof. By \Cref{clm:euclidean-kmeans-cluster-mass}, with probability at
least \(1-\delta/2\), we have
\(|w_q-\widehat w_q|\lesssim \eps^2\widehat w_q\) for every \(q\in Q\).
Together with \Cref{eq:euclidean-kmeans-g-range}, this gives, for every
\(j\in[k]\),
\[
\left|
\sum_{q\in Q_j}(w_q-\widehat w_q)g_{C,\beta}(q)
\right|
\lesssim
\eps\sum_{q\in Q_j}\widehat w_q r_q^2 .
\]
Moreover, since \(\pi_j(q)\ge r_q^2/(2\Phi_j)\), we have
\(\widehat w_q r_q^2\le 2\Phi_j/m_j\) for every \(q\in Q_j\), and hence
\(\sum_{q\in Q_j}\widehat w_q r_q^2\le 2\Phi_j\). Summing over
\(j\in[k]\) gives
\[
\left|
\sum_{q\in Q}(w_q-\widehat w_q)g_{C,\beta}(q)
\right|
\lesssim
\eps\sum_{j\in[k]}\Phi_j
=
\eps\Wtwo .
\]
\end{proof}

\section{Acknowledgements} Milind Prabhu was supported by the NSF grant CCF-2327011. 
Chris Schwiegelshohn and Sudarshan Shyam were supported in part by a Google Research Award.  Erik Waingarten was supported by the NSF grant CCF-2337993.

\bibliographystyle{alpha}
\bibliography{ref}

\appendix
\section{Related Work}
\label{sec:related}
The first coreset definition as it is nowadays understood for clustering problems was given by Har-Peled and Mazumdar \cite{HaM04}. Initially, the techniques were mainly focused on low-dimensional Euclidean spaces \cite{FrahlS2005,HaK07}, and are still an active area of research today \cite{AfshaniS24,HuangHH023}. Chen's seminal paper \cite{Chen09} brought sampling and learning theory to coresets, initially mostly via uniform sampling. Feldman, Langberg, and Schulman \cite{FL11,LS10} then introduced  the sensitivity sampling distribution, arguably the definitive algorithm in this line of work. Thereafter, for almost a decade, most coreset papers either focused on dimension reduction techniques for high dimensional Euclidean spaces \cite{BecchettiBC0S19,CharikaW25,FeldmanSS20,MakarychevMR19,SohlerW18,FengKW21}, or by applying the sensitivity sampling framework to other metrics such as doubling metrics \cite{HuangJLW18}, graph metrics \cite{BravermanJKW21,BakerBHJK020}, or metrics on time series \cite{ConradiKPR24,HuangSV21,munteanu2026terminal}.
Cohen-Addad, Saulpic, and Schwiegelshohn \cite{CSS21} proposed the group sampling algorithm, which ultimately yielded optimal coreset bounds for most of these problems \cite{Cohen-AddadD0SS25,CGSS22,Cohen-AddadLSSS22,HLW23}. 
It later turned out that sensitivity sampling is likewise optimal \cite{sensitivitycoresets,prabhu2026approximation}, empirically preferable \cite{SchwiegelshohnS22}, fault tolerant \cite{PSS26b}, and can be implemented extremely efficiently \cite{CWWZ25,HenzingerK20,schwiegelshohnsettling} in various settings.

\section{Concentration Inequalities}


We first state two standard concentration inequalities. 


\begin{lemma}[Hoeffding's inequality {\cite[Theorem~2.2.6]{vershynin2026high}}]
\label{lem:hoeffding}
Let \(Z_1,\ldots,Z_m\) be independent mean-zero random variables. Suppose that
there exists a constant \(M>0\) such that 
\(
|Z_j|\le M
\text{ almost surely for all } j\in[m].
\)
Then, for every \(\delta\in(0,1)\), with probability at least \(1-\delta\),
\[
\left|\sum_{j=1}^m Z_j\right|
\le
\sqrt{
2mM^2
\log\frac{2}{\delta}
}.
\]
\end{lemma}

\begin{lemma}[Bernstein's inequality {\cite[Theorem~2.9.5]{vershynin2026high}}]
\label{lem:bernstein}
Let \(Z_1,\ldots,Z_m\) be independent mean-zero random variables. Suppose
there exists a constant \(M>0\) such that \(|Z_j|\le M\) almost surely for
all \(j\in[m]\), and define
\[
V=\sum_{j=1}^m \mathbb E[Z_j^2].
\]
Then, for every \(\delta\in(0,1)\), with probability at least \(1-\delta\),
\[
\left|\sum_{j=1}^m Z_j\right|
\lesssim
\sqrt{V\log\frac{2}{\delta}}
+
M\log\frac{2}{\delta}.
\]
\end{lemma}

\subsection{Concentration via Pseudo-dimension}
\label{app:pseudodimension-tools}

In the Euclidean sections, we need uniform concentration for families of
normalized sampled summands indexed by the centers and dual variables. We
use a standard two-step approach: first bound the pseudo-dimension of the
relevant function class, and then apply a uniform convergence bound for
bounded real-valued functions. The pseudo-dimension bounds are obtained by
showing that each threshold comparison is determined by the signs of a
bounded number of low-degree polynomials, and then applying Warren's
theorem. This appendix collects the common tools used in both the
Euclidean \(\kmedian\) and Euclidean \(\kmeans\) analyses.

\begin{definition}[Pseudo-dimension]
Let \(\mathcal F\) be a class of real-valued functions on a domain
\(\mathcal X\). A set \(S=\{x_1,\ldots,x_m\}\subseteq\mathcal X\) is
pseudo-shattered by \(\mathcal F\) if there are thresholds
\(\tau_1,\ldots,\tau_m\in\mathbb R\) such that, for every
\(I\subseteq[m]\), there is \(f\in\mathcal F\) satisfying
\[
f(x_i)\ge \tau_i \quad\text{for } i\in I,
\qquad
f(x_i)<\tau_i \quad\text{for } i\notin I .
\]
The pseudo-dimension of \(\mathcal F\), denoted \(\pdim(\mathcal F)\), is
the largest size of a set pseudo-shattered by \(\mathcal F\).
\end{definition}

The following lemma is a uniform convergence bound for bounded
real-valued classes of finite pseudo-dimension; see, for ex.,
\cite[Chapter 11]{AnthonyBartlett1999} and
\cite[Chapter 2.6]{vdVW1996}.

\begin{restatable}[Uniform convergence]{lemma}{UniformConvergence}
\label{lem:uniform-convergence}
Let \(\mathcal F\) be a class of real-valued functions with
\(\pdim(\mathcal F)\le D\). Let \(Z_1,\ldots,Z_m\) be independent random
variables, not necessarily identically distributed. Suppose that every
\(f\in\mathcal F\) satisfies \(|f(Z_t)|\le 1\) for every \(t\). Then, with
probability at least \(1-\delta\),
\[
\sup_{f\in\mathcal F}
\left|
\frac1m\sum_{t=1}^m f(Z_t)
-
\frac1m\sum_{t=1}^m \mathbb E f(Z_t)
\right|
\lesssim
\sqrt{\frac{D\log m+\log(1/\delta)}{m}}.
\]
\end{restatable}

Bounding the pseudo-dimension of a function class directly is sometimes
challenging.  We will use Warren's theorem as a standard tool for this purpose.  It applies when the relevant threshold comparisons can be
described by low-degree polynomial inequalities. For a real number \(x\), let \(\operatorname{sign}(x)=1\) if \(x\ge 0\)
and \(\operatorname{sign}(x)=-1\) otherwise. 
\begin{lemma}[Warren's theorem {\cite{warren1968lower}}]
\label{thm:warren}
Let \(P_1,\ldots,P_M\) be real polynomials in \(s\) variables, each of
degree at most \(\Delta\). If \(M\ge s\), then the number of sign patterns
\((\operatorname{sign}(P_1(x)),\ldots,\operatorname{sign}(P_M(x)))\), over
all \(x\in\mathbb R^s\), is at most
\(\left(4e\Delta M/s\right)^s\).
\end{lemma}

\section{Omitted Proofs}



\subsection{Reduction to Upper Capacities}

We also briefly outline how coresets for clustering with exact demands implies coresets for upper bounded capacitated clustering. Similar reductions apply to
cardinality lower bounds and fair clustering. These reductions are folklore,
but we include the argument for completeness and future reference.

\begin{proposition}[Reduction from exact demands to upper capacities]
\label{prop:reduction}
Let \(P\) be a point set, and let \(u=(u_1,\ldots,u_k)\) be an
upper-capacity vector. Suppose that \(Q\) is an \(\varepsilon\)-coreset for
\(P\) with respect to exact demand vectors \(h\), meaning that for every set
of centers \(C\) and every feasible demand vector \(h\),
\[
(1-\varepsilon)\cost_P(C,h)
\le
\cost_Q(C,h)
\le
(1+\varepsilon)\cost_P(C,h).
\]
Then \(Q\) is an \(\varepsilon\)-coreset for \(P\) with respect to the
upper-capacity constraints
\[
|C_i|\le u_i
\qquad\text{for all } i\in\{1,\ldots,k\}.
\]
\end{proposition}

\begin{proof}
Fix a set of centers \(C\). The capacitated cost on \(P\) can be written as
a minimization over exact demand vectors:
\[
\cost_P^{\mathrm{cap}}(C,u)
=
\min_{\substack{h:\, \sum_i h_i = |P| \\ 0\le h_i\le u_i}}
\cost_P(C,h).
\]
Similarly,
\[
\cost_Q^{\mathrm{cap}}(C,u)
=
\min_{\substack{h:\, \sum_i h_i = |P| \\ 0\le h_i\le u_i}}
\cost_Q(C,h).
\]
Let \(h_P\) be an exact demand vector attaining the minimum for \(P\), and
let \(h_Q\) be one attaining the minimum for \(Q\). Then
\[
\cost_Q^{\mathrm{cap}}(C,u)
=
\cost_Q(C,h_Q)
\le
\cost_Q(C,h_P)
\le
(1+\varepsilon)\cost_P(C,h_P)
=
(1+\varepsilon)\cost_P^{\mathrm{cap}}(C,u).
\]
For the other direction,
\[
\cost_Q^{\mathrm{cap}}(C,u)
=
\cost_Q(C,h_Q)
\ge
(1-\varepsilon)\cost_P(C,h_Q)
\ge
(1-\varepsilon)\cost_P^{\mathrm{cap}}(C,u),
\]
where the last inequality uses the fact that \(h_Q\) is feasible for the
upper-capacitated problem on \(P\). Therefore,
\[
(1-\varepsilon)\cost_P^{\mathrm{cap}}(C,u)
\le
\cost_Q^{\mathrm{cap}}(C,u)
\le
(1+\varepsilon)\cost_P^{\mathrm{cap}}(C,u),
\]
as desired.
\end{proof}

\subsection{Dimension-Independent Coreset Bounds}
\label{sec:dimension-independent-bound}

In \Cref{lem:euclidean-kmedian-dimension-dependent} and
\Cref{lem:euclidean-kmeans-dimension-dependent}, we proved
dimension-dependent coreset bounds for Euclidean capacitated \(\kmedian\)
and \(\kmeans\). We now explain how to remove the dependence on the
ambient dimension. The reduction has two standard ingredients. First, we
use an existing Euclidean coreset construction to reduce the support size
to \(\poly(k/\eps)\). Second, once the support is small, we use a terminal
embedding to reduce the dimension to \(\widetilde O(\eps^{-2})\).

We use the small sized coresets as a black box.

\begin{theorem}[Polynomial-size Euclidean coresets~\cite{BravermanCJKST022,Huang0L025}]
\label{thm:preliminary-euclidean-coreset}
For Euclidean capacitated \(\kmedian\) and \(\kmeans\), every instance
\(P\subseteq\mathbb R^d\) admits an \(O(\eps)\)-coreset
\(S\subseteq P\) of size
\(
|S|=\poly(k/\eps).
\)
\end{theorem}

We first replace the input points \(P\) by
\(S\), reducing the support size to  \(\poly(k/\eps)\). Since coresets
compose, it is enough to construct an \(O(\eps)\)-coreset for \(S\). 

Here \(S\) is a weighted instance. Although the Euclidean
dimension-dependent bounds were stated for unweighted inputs to avoid
extra notation, the proofs extend verbatim to weighted inputs: replace
cardinalities by total weights, replace sums by weighted sums, and perform
the same cluster-wise rescaling with respect to the total weight of each
reference cluster.

We next recall terminal embeddings. Unlike the Johnson--Lindenstrauss
lemma, which preserves pairwise distances inside a finite point set,
terminal embeddings preserve distances from a fixed finite terminal set to
all query points in the ambient space.

\begin{definition}[Terminal embedding]
Let \(T\subseteq\mathbb R^d\) be a finite terminal set. A map
\(\varphi:\mathbb R^d\to\mathbb R^{d'}\) is a terminal embedding for
\(T\) with distortion \(1+\eta\) if, for every terminal \(t\in T\) and
every query point \(y\in\mathbb R^d\),
\[
\|t-y\|
\le
\|\varphi(t)-\varphi(y)\|
\le
(1+\eta)\|t-y\|.
\]
\end{definition}

We use the following optimal terminal-embedding bound of Narayanan and
Nelson \cite{narayanan2019optimal}.

\begin{theorem}
\label{thm:terminal-embedding}
Let \(T\subseteq\mathbb R^d\) be a finite set of terminals. For every
\(\eta\in(0,1)\), there is a terminal embedding
\(\varphi:\mathbb R^d\to\mathbb R^{d'}\) for \(T\) with distortion
\(1+\eta\), where
\[
d'=O(\eta^{-2}\log |T|).
\]
\end{theorem}

This is exactly the type of dimension reduction needed for clustering:
the data points are the terminals, while the centers are arbitrary query
points.

\begin{lemma}
\label{lem:terminal-transfer-exact-loads}
Let \(X\subseteq\mathbb R^d\) be a weighted instance, and let
\(\varphi\) be a terminal embedding for the support of \(X\) with
distortion \(1+\eta\). Then, for every center tuple \(C\) and every 
\loadvector \(h \in n \Delta_k\),
\[
\cost_{\varphi(X)}^{\kmedian}(\varphi(C),h)
=
(1\pm O(\eta))\cost_X^{\kmedian}(C,h),
\]
and
\[
\cost_{\varphi(X)}^{\kmeans}(\varphi(C),h)
=
(1\pm O(\eta))\cost_X^{\kmeans}(C,h).
\]
\end{lemma}

\begin{proof}
For \(\kmedian\), every point-center distance is preserved up to a
\(1\pm O(\eta)\) factor. Hence the cost of every feasible exact-load
assignment is preserved up to the same factor, and taking the minimum over
assignments preserves the guarantee.

For \(\kmeans\), the same distance preservation gives squared-distance
preservation up to a \(1\pm O(\eta)\) factor, after adjusting constants.
The same assignment-wise argument applies.
\end{proof}

\paragraph{Applying the reduction.}
For \(\kmedian\), first replace \(P\) by the preliminary
\(O(\eps)\)-coreset \(S\subseteq P\) of size \(\poly(k/\eps)\). Apply
\Cref{thm:terminal-embedding} to the support of \(S\) with
\(\eta=\Theta(\eps)\), obtaining dimension
\[
d'=O(\eps^{-2}\log |S|)=\widetilde O(\eps^{-2}).
\]
By \Cref{lem:terminal-transfer-exact-loads}, a coreset for the embedded
instance can be pulled back to a coreset for \(S\). Composing with the
preliminary coreset then gives a coreset for \(P\). Applying
\Cref{lem:euclidean-kmedian-dimension-dependent} in dimension \(d'\)
therefore gives size
\[
\widetilde O\!\left(kd'\eps^{-2}+k\eps^{-2}\log(1/\delta)\right)
=
\widetilde O\!\left(k\eps^{-4}+k\eps^{-2}\log(1/\delta)\right),
\]
as claimed in \Cref{thm:euclidean-kmedian}. The three approximation
losses are each set to a sufficiently small constant multiple of \(\eps\),
so they compose to \(1\pm\eps\).

The proof for \(\kmeans\) is identical, using
\Cref{lem:euclidean-kmeans-dimension-dependent} instead; this gives size
\[
\widetilde O\!\left(k\eps^{-6}+k\eps^{-4}\log(1/\delta)\right),
\]
as claimed in \Cref{thm:euclidean-kmeans}.

\end{document}